\documentclass[aps,pre,twocolumn,superscriptaddress]{revtex4-2}
\usepackage[utf8]{inputenc}
\usepackage{graphicx}
\usepackage{epstopdf}
\usepackage{bm}
\usepackage{amsmath}
\usepackage{verbatim}
\usepackage{amssymb}
\usepackage{listings}
\usepackage{color}
\definecolor{darkblue}{rgb}{0,0,0.6}
\usepackage{here}
\usepackage[
    colorlinks,
    linkcolor=darkblue,
    citecolor=darkblue,
    urlcolor=darkblue
]{hyperref}
\usepackage{mathtools}
\DeclareRobustCommand{\average}[1]{\left\langle #1 \right\rangle}
\renewcommand{\vec}[1]{\boldsymbol{#1}}

\newcommand{\fig}[1]{Fig.~\ref{#1}}
\newcommand{\Sec}[1]{Sec.~\ref{#1}}

\newcommand{\eq}[1]{Eq.~(\ref{#1})}

\graphicspath{{figure/}}

\begin{document}
    \title{Fast free-energy estimation for complex systems with nonreversible simulated tempering}
    \author{Yoshihiko Nishikawa}
    \affiliation{Department of Physics, Nagoya University, Nagoya 464-8602, Japan}

    \author{Koji Hukushima}
    \affiliation{Department of Basic Science, Graduate School of Arts and Sciences, The University of Tokyo, 3-8-1 Komaba, Meguro, Tokyo 153-8902, Japan}

\begin{abstract}
    We discuss free-energy estimation for complex systems using simulated tempering with Wang--Landau-type feedback. We show that introducing nonreversible dynamics in temperature space and multiple temperature updates per step significantly accelerates the convergence of the free-energy estimate. We then apply the algorithm to the three-dimensional Edwards-Anderson spin-glass model. Our results show that the algorithm can precisely estimate the free energy up to a large system size $L \simeq 64$, and that subsequent equilibrium simulated tempering simulations with the estimated free energy can rapidly sample from the Boltzmann distributions over a wide temperature range that covers the spin-glass transition temperature. The convergence time of our algorithm grows much more slowly with system size than that of another known algorithm for on-the-fly free-energy estimation. 
\end{abstract}

    \maketitle

    \section{Introduction}
    Complex systems such as glasses, liquid crystals, biomolecular systems, and frustrated spin systems often exhibit slow physical dynamics, particularly at low temperatures and high densities. These slow dynamics provide important insights into the underlying thermodynamic behaviors. Nevertheless, they pose a serious challenge for numerical simulations, since reaching equilibrium and precisely estimating physical quantities of such systems require high computational costs. Advanced Monte Carlo algorithms can overcome the slow dynamics of some systems by implementing `unphysical' dynamics while satisfying detailed balance. Cluster algorithms \cite{Swendsen1987nonuniversal,Wolff1989collective,Dress1996cluster,Liu2004rejection} are the earliest successful examples in the literature. These algorithms strongly reduce critical slowing down in the vicinity of ferromagnetic and liquid-gas critical points, although significant acceleration is achieved only for a very limited class of systems. Another family of efficient algorithms includes generalized-ensemble and extended-ensemble algorithms~\cite{Iba_2001}. These methods were developed extensively since the 1990s and are now standard in a wide range of fields, from particle physics and condensed matter to biophysics and machine learning.

    Simulated tempering (ST) \cite{Marinari1992,Lyubartsev1992} is one of the earliest extended-ensemble algorithms. The basic idea of the algorithm is simple: Instead of fixing the temperature and waiting for the system to reach equilibrium and decorrelate, the algorithm dynamically changes the temperature while preserving the correct equilibrium distribution, thereby bypassing the slow dynamical pathways at low temperatures. This is achieved by adjusting the temperature according to a probability that satisfies the balance condition. By bringing the system to sufficiently high temperatures, where decorrelation quickly takes place, ST yields completely independent \textit{trips} from the highest temperature to the lowest temperature. Each trip thus gives an independent configuration at the lowest temperature. The low-temperature correlation timescale is thus controlled by the round-trip time in the temperature space. As these trips are independent, parallel ST simulations that require no communication with each other are possible, efficiently providing independent configurations at the lowest temperature. These \textit{regenerative} features \cite{Mykland1995regeneration} of ST match the massively parallel architectures of modern computers but have only recently been discussed in the statistics community \cite{Biron-Lattes2025automatic}. 

    Although these features are attractive, ST requires careful tuning of a parameter which we call a bias function \cite{Marinari1992}. This bias function directly controls the distribution and the round-trip timescale in the temperature space. An optimal choice for the bias function is the free energy multiplied by the inverse temperature, which yields a uniform distribution over all the temperatures. However, the free energy of a given system is generally not known a priori, and its estimation is a daunting task especially for complex and large systems. This drawback of ST was already pointed out in the original paper by Marinari and Parisi \cite{Marinari1992}, and has been a major obstacle to the practical use of ST. Exchange Monte Carlo method~\cite{Geyer1991, Hukushima1996}, also known as parallel tempering (PT)~\cite{Marinari1998}, on the other hand, does not require prior estimates of the free energy, which has contributed to its wider use in practice than ST. 
    
    Nevertheless, some studies \cite{Mitsutake2000replica, Mitsutake2004replica, Park2007choosing, Nguyen2013simulated, Sakai2016learning} have discussed free-energy estimation in the context of ST. For instance, Refs.~\cite{Mitsutake2000replica, Mitsutake2004replica} use preliminary PT runs and the reweighting technique \cite{Ferrenberg1989optimized, Munger1991reweighting} to estimate the free energy for subsequent ST runs. This approach, however, requires multiple stages of simulations and statistical analyses before production runs. Also, once sufficiently equilibrated samples have been obtained from the preliminary PT runs, switching to ST offers only little practical benefit compared with continuing the PT simulation. Refs.~\cite{Park2007choosing,Nguyen2013simulated,Sakai2016learning}, on the other hand, adopt a different strategy, where the bias function is adaptively updated on the fly during a ST simulation through simple thermodynamic integration, allowing a seamless transition to a ST production run. Although successful for some small systems, this becomes impractical for complex systems, as the temperature is effectively fixed at the given lowest temperature from the outset, and the system needs to equilibrate there without the aid of ST. 
    Consequently, the timescale needed to obtain a precise free-energy estimate is controlled by the equilibration timescale at the lowest temperature, which can be very long.
    
    Here, we present an improved ST-like algorithm for efficient free-energy estimation. As in the conventional ST, the temperature fluctuates with time, but the bias function is iteratively updated to flatten the temperature distribution, eventually yielding a precise estimate of the free energy. This approach can be viewed as a generalization of the Wang--Landau algorithm \cite{Wang2001, WangLandau2001} and was first studied in Ref.~\cite{Zhang2007simulation}. We incorporate a simple lifted nonreversible dynamical rule into temperature updates of the algorithm to accelerate convergence. Nonreversible Markov-chain dynamics that converges to the equilibrium distribution have been actively studied over the past three decades and outperform reversible algorithms in a wide range of systems \cite{chen1999lifting,diaconis2000analysis,Bernard2009, Suwa2010markov,Turitsyn2011irreversible,Fernandes2011nonreversible,Engel2013harddisk, Michel2014generalized,nishikawa2015event,Isobe2015hardsphere,Lei2018irreversible, Lei2019eventchain,Klement2019efficient,Maggs2024nonreversible,Li2022harddisk, Ghimenti2024irreversible,Berthier2024MonteCarlo,Nishikawa2025irreversible}. Nonreversible variants of ST have also been studied in recent works \cite{Sakai2016irreversible,Faizi2020simulated,Biron-Lattes2025automatic}, where These studies have mainly focused on accelerating temperature-space mixing and equilibrium sampling, whereas the use of nonreversible dynamics for adaptive free-energy estimation for more practical and complex systems remains largely unexplored.

    We further improve the efficiency of the algorithm by increasing the number of temperature updates in each tempering process. This simple change largely accelerates the convergence of the algorithm. We demonstrate this by applying the algorithm to the three-dimensional Edwards-Anderson spin glass model, a representative complex model with slow dynamics at low temperatures. We also quantify the convergence timescale of the algorithm, and show that it outperforms the aforementioned on-the-fly free-energy estimation algorithm \cite{Nguyen2013simulated} at large system sizes. These results establish our algorithm as a practical approach for estimating the free energy of complex systems with large sizes. 

    The rest of the paper is organized as follows. In \Sec{sec: background}, we review the simulated tempering and the original Wang--Landau algorithms, which constitute our algorithm we present in \Sec{sec: WLST}. We then discuss our numerical results of the algorithm on the spin-glass model in \Sec{sec: numerical results on EASG model}. \Sec{sec: conclusion} summarizes the paper.

    \section{Algorithmic background}
    \label{sec: background}

    In this section, we review simulated tempering and the original Wang--Landau algorithm formulated in the energy space, which provide the two main ingredients of our algorithm for efficient free-energy estimation. For simulated tempering, we explain the need to tune the bias function for the practical use of ST. For the Wang--Landau algorithm, we briefly discuss its generalization to other quantities.
   
    \subsection{Simulated tempering}
    \label{sec: simulated tempering}
    Simulated tempering extends the state space by allowing temperature, or any other intensive parameters, to fluctuate during the simulation. Here, we let temperature take any one value in a discrete set of inverse temperatures $\{\beta_i\}$. The extended partition function of the system with the fluctuating temperature is then 
    \begin{align}
        \mathcal Z
        = \sum_i \text{Tr}_{\vec \sigma} \exp\left( -\beta_i H(\vec \sigma) + g(\beta_i)\right),
    \end{align}
    where $\vec \sigma$ denotes the microscopic degrees of freedom of the system, $H$ is the Hamiltonian, and $g$ is the bias function.
    The normalized joint distribution of $\vec \sigma$ and $\beta_i$ is 
      \begin{align}
        \pi(\vec \sigma, \beta_i) = \frac{1}{\mathcal Z} \exp(-\beta_i H(\vec \sigma) + g(\beta_i)).
        \label{eq: joint dist}
    \end{align}

    An ST simulation consists of two distinct processes: The usual updating process for the microscopic degrees of freedom $\vec \sigma$ at fixed $\beta$ and the tempering process that changes the temperature with fixed $\vec \sigma$.
    The latter must leave the joint distribution in Eq.~(\ref{eq: joint dist}) invariant. This can be realized by using the Metropolis transition probability 
    \begin{align}
        P(\beta_i \to \beta_j | \vec \sigma) = \min\left[ 1, \frac{\exp( - \beta_j H(\vec \sigma) + g(\beta_j))}{\exp(-\beta_i  H(\vec \sigma) + g(\beta_i))}\right].
        \label{eq:Met transition prob}
    \end{align}
    The marginal stationary distribution of $\beta$ is controlled by the bias function $g(\beta)$: 
    \begin{align}
        \begin{split}
        \pi(\beta) &= \frac{\text{Tr}_{\vec \sigma} \exp\left(-\beta H(\vec \sigma) + g(\beta)\right)}{\mathcal Z}\\
        &= \frac{\exp\left(-\varphi(\beta) + g(\beta)\right)}{\mathcal Z},
        \end{split}
        \label{eq: beta dist for ST}
    \end{align}
   where $\varphi(\beta) = \beta F(\beta) = -\log Z(\beta)$ is the dimensionless free energy of the system at inverse temperature $\beta$. We will refer to $\varphi(\beta)$ as the free energy for simplicity in the following. To sample all of the prescribed temperatures with the same probability, the bias function must be chosen as $g(\beta) = \varphi(\beta)$. However, as the free energy of systems is generally not known a priori, this requires preliminary runs to estimate the free energy, which can be computationally costly, especially for complex systems with slow dynamics at low temperatures. 
   This additional tuning requirement is a practical disadvantage of ST compared with PT. An efficient adaptive strategy for learning the bias function that encourages exploration over the entire temperature range, including the lowest temperatures, is therefore highly desirable, particularly one that does not require prior equilibration at the lowest temperatures.

    \subsection{Wang--Landau algorithm in energy space}
    \label{sec: Wang--Landau}
    
    The Wang--Landau (WL) algorithm \cite{Wang2001, WangLandau2001} is a Monte Carlo algorithm for estimating the density of states (DOS) as a function of energy,  
    \begin{align}
        \Omega(E) = \text{Tr}_{\vec \sigma} \delta(H(\vec \sigma) - E),
    \end{align}
    up to an overall multiplicative constant. The transition probability evolves over time and depends on the entire history of the simulation. The dynamics of the algorithm is thus non-Markovian. For an instantaneous estimate $\tilde\Omega(E)$, the microscopic degrees of freedom are updated using a transition probability that has the stationary distribution proportional to $\hat \Omega^{-1}(E)$. As the magnitude of the modifications to $\hat \Omega(E)$ is progressively reduced, the estimate eventually converges to $\Omega(E)$. 

   Compared to the canonical simulation, the WL algorithm needs three additional ingredients: the DOS estimate $\hat \Omega(E)$, a histogram $N(E)$ that records the number of visits to each energy level, and a modification parameter $f$. A WL simulation is typically initialized with $\hat \Omega(E) = 1$, $N(E) = 0$, and $f = e$. A proposed change of the system state from $\vec \sigma$ to $\vec \sigma^\prime$ is accepted with probability 
    \begin{align}
        P(\vec \sigma \to \vec \sigma^\prime) 
        = \min\left[ 1, \frac{\hat \Omega(H(\vec \sigma))}{\hat \Omega(H(\vec \sigma^\prime))} \right].
    \end{align}
    After each attempted update, the  DOS estimate $\hat \Omega(E)$ and the histogram $N(E)$ at the instantaneous $E$ are then updated as 
    \begin{align}
        \begin{split}
        N(E) &\leftarrow N(E) + 1,\\
        \log \hat \Omega(E) &\leftarrow \log \hat \Omega(E) + \log f.
        \end{split}
        \label{eq: WL rule}
    \end{align}
    Once the histogram is sufficiently `flat', all entries of $N(E)$ are reset to zero, the modification factor $f$ is reduced as $f\leftarrow \sqrt{f}$. The simulation stops when $f$ reaches down to a given threshold.

    Whereas convergence to the exact DOS of the WL algorithm has been proved \cite{Zhou2005,Fort2015}, the convergence speed and precision of the estimate depend on the implementation details of the algorithm. These include the criterion to judge whether the histogram is sufficiently flat, the rule for reducing the parameter $f$, and, most importantly, the terminal value of $f$. A very small terminal value for $f$, typically $\log f \approx 10^{-8}$, yields a high-precision estimate of $\Omega(E)$ \cite{Wang2001}. However, the WL simulation by itself cannot compute the equilibrium value for physical quantities not directly related to $E$, and a subsequent {multicanonical} run \cite{Berg1991, Berg1992} using the approximate DOS obtained from the WL simulation is needed to estimate these canonical averages. Although the performance of the multicanonical simulation depends on the precision of the DOS, the exact DOS is not required. As long as the system can explore the full energy range during a multicanonical run, one can compute the canonical averages using the reweighting technique. If a more precise estimate for the DOS is needed, it can be refined using the histogram of energy $N_\text{muca}(E)$ obtained from the multicanonical run as
    \begin{align}
        \tilde{\Omega}_{\mathrm{new}}(E) = \tilde{\Omega}_{\mathrm{old}}(E) N_\text{muca}(E).
    \end{align}
    This refinement can be repeated by modifying the DOS estimate in a subsequent multicanonical simulation. 

    The WL algorithm can also be extended to other ensembles, to estimate the DOS for an arbitrary quantity $A(\vec \sigma)$ depending on the system configuration \cite{Hansmann1993MultimagneticalSimulations, Zhou2005, Zhou2006wanglandau, Tsai2009, Atchade2010wanglandau, Nishikawa2020lattice}
    \begin{align}
        \Omega_{\beta}(A) = \text{Tr}_{\vec \sigma} \delta(A - A(\vec \sigma))\exp(-\beta H(\vec \sigma)).
        \label{eq: dos for A}
    \end{align}
    At a fixed inverse temperature $\beta$, the acceptance probability is modified as
    \begin{align}
        P(\vec \sigma \to \vec \sigma^\prime) 
        = \min\left[ 1, e^{-\beta(H(\vec \sigma^\prime) - H(\vec \sigma))} 
        \frac{\hat \Omega_\beta(A(\vec \sigma))}{\hat \Omega_\beta(A(\vec \sigma^\prime))} \right],
    \end{align}
    with $\hat \Omega_\beta(A)$ the instantaneous estimate of $\Omega_{\beta}(A)$.

    \section{Free-energy estimation using simulated tempering}
    \label{sec: WLST}

    By leveraging the flexibility of the WL algorithm, we further generalize if for free-energy estimation. Introducing the extended state space $\tilde{\vec \sigma} = (\vec \sigma, \beta)$, we rewrite the canonical partition function at $\beta^\prime$ as
    \begin{align}
        Z(\beta^\prime)= \text{Tr}_{\tilde{\vec \sigma}} \delta(\beta - \beta^\prime) \exp(-\beta H(\vec \sigma)).
    \end{align}
    A comparison with \eq{eq: dos for A} implies that we can estimate $Z(\beta)$ and the free energy as a function of temperature, in essentially the same way as in the original WL algorithm \cite{Zhang2007simulation, Atchade2010wanglandau}: During the tempering process, the histogram $N(\beta)$ and the instantaneous estimate for the free energy are updated using a feedback protocol analogous to that in \eq{eq: WL rule}. With decreasing $f$, the free-energy estimate becomes gradually more precise and eventually converges to the exact one at $f \to 1$, up to an additive constant. By setting the bias function to the estimated free-energy, one can perform equilibrium ST simulations to measure the canonical averages, with an approximately uniform temperature distribution $\pi(\beta)$. This is exactly in parallel with the original WL simulation followed by multicanonical simulations. In the following, we refer to this as the WLST algorithm.

    In the conventional implementation of ST \cite{Marinari1992}, the number of temperature updates per tempering step, denoted by $\ell$, is often set to $1$. When the temperature is reversibly updated between neighboring values, the resulting dynamics in temperature space is slow and diffusive. This could be easily improved by incorporating a nonreversible dynamical rule for the temperature updates, as demonstrated in Refs.~\cite{Sakai2016irreversible,Faizi2020simulated,Biron-Lattes2025automatic}. These studies, however, still set $\ell = 1$, although increasing $\ell$ should further accelerate mixing in temperature space and convergence of the WLST algorithm. In what follows, we thus set $\ell = M / 2$, where $M$ is the number of temperatures. Other choices for $\ell$ with a weaker dependence on $M$, such as $\ell \sim M^{1/2}$, could provide comparable performance. This large value of $\ell$ increases the computational cost of each tempering step compared to the conventional ST implementation with $\ell=1$. Nevertheless, since $M$ is usually much smaller than the number of microscopic degrees of freedom, the additional cost of performing $O(M)$ temperature updates remains smaller than that of updating the microscopic configuration. Note that the nonreversible Gibbs sampler for $\beta$ discussed in Ref.~\cite{Faizi2020simulated} also takes $O(M)$ computations for building a list of the transition probabilities.

    We now describe the nonreversible temperature dynamics used in our algorithm. Suppose that the current inverse temperature is set to $\beta_i \in \{\beta_0, \cdots, \beta_{M-1}\}$, where $\beta_0 < \beta_1 < \cdots < \beta_{M-1}$. We introduce a \textit{lifting} variable $v \in \{\pm 1\}$ that specifies the direction of motion in temperature space. For $(\vec \sigma, \beta_i, v)$, we propose a change from $\beta_i$ to $\beta_{i+v}$ with probability one and accept it with the Metropolis probability \eq{eq:Met transition prob}. If the temperature change is accepted, we set $\beta_i \gets \beta_{i+v}$ and make the next proposal from $\beta_{i+v}$ to $\beta_{i+2v}$ in the same direction. We repeat this process until a change is rejected, at which point a lifting event reverses the direction as $v \gets -v$. The subsequent $\beta$ updates follow with the updated $v$. If the system reaches the maximum or minimum $\beta$s, i.e. when $\beta = \beta_0$ or $\beta = \beta_{M-1}$, an attempted proposal is rejected with probability one and $v$ is flipped so that $\beta$ goes backward. The tempering process terminates when the total number of attempts reaches $\ell$.

    This nonreversible, lifted dynamics was designed carefully to preserve the target joint distribution (\eq{eq: joint dist}), upon marginalizing its stationary distribution over the lifting variable $v$, when the bias function is fixed. To prove that, let $\tilde{\pi}(\vec \sigma, \beta, v) = \pi(\vec \sigma, \beta) / 2$ denote the stationary distribution on the lifted state space $(\vec \sigma, \beta, v)$. We then compare the total incoming and outgoing probability flows for each lifted state. For a fixed configuration $\vec \sigma$, the total outgoing flow from $(\vec \sigma, \beta_i, v)$ is 
    \begin{align}
        J_{\mathrm{out}}(\vec{\sigma},\beta_i,v) 
        &= \tilde{\pi}(\vec{\sigma},\beta_i,v) P(\beta_i\to\beta_{i+v}\mid\vec{\sigma}) \nonumber \\
        & \quad + \tilde{\pi}(\vec{\sigma},\beta_i,v)\left(
        1-P(\beta_i\to\beta_{i+v}\mid\vec{\sigma})
        \right),\nonumber \\
        &= \tilde{\pi}(\vec{\sigma},\beta_i,v), 
    \end{align}
    where $P(\beta \to \beta^\prime\mid\vec\sigma)$ is given in \eq{eq:Met transition prob}. The total incoming flow, on the other hand, is 
    \begin{align}
       J_{\mathrm{in}}(\vec \sigma, \beta_i,v) 
        &= \tilde{\pi}(\vec \sigma, \beta_{i-v}, v) P(\beta_{i-v} \to \beta_i\mid \vec \sigma) \nonumber \\
        & \quad + \tilde{\pi}(\vec \sigma, \beta_i, -v) \left(1 - P(\beta_i \to \beta_{i-v}\mid\vec\sigma)\right).
    \end{align}
    These two probability flows are identical, as the Metropolis acceptance probability in Eq.~(\ref{eq:Met transition prob}) satisfies the pairwise identity 
    \begin{align}
        &\tilde \pi(\vec{\sigma},\beta_{i-v}, v)P(\beta_{i-v}\to\beta_i\mid\vec{\sigma}) \nonumber \\
        &\quad = 
        \tilde \pi(\vec{\sigma},\beta_i, -v) P(\beta_i\to\beta_{i-v}\mid\vec{\sigma}). 
        \label{eq:pairwise_identity}
    \end{align} 
    The same stationarity condition holds at the endpoints, i.e., $\beta_0$ and $\beta_{M-1}$ when an attempted move outside the temperature range is treated as a rejection with probability one. Thus, the stationary distribution of the lifted dynamics is indeed $\tilde \pi$. Eventually, marginalizing $\tilde \pi$ over the lifting variable yields the joint distribution of \eq{eq: joint dist}. Note that \eq{eq:pairwise_identity} is an equation for the probability flows of different lifting variables and does not imply detailed balance for the lifted dynamics, since a transition from $(\vec\sigma,\beta_{i+v},v)$ to $(\vec\sigma,\beta_i,v)$ never occurs in the dynamics.

    In our algorithm, we modify the temperature histogram and the free-energy estimate at each of $\ell$ temperature-update attempts. Each tempering step therefore accumulates $\ell$ counts in total to the histogram. For $M \ll N$, the computational cost of performing $O(M)$ temperature updates remains modest compared to that of updating the $N$ microscopic degrees of freedom, while the increased frequency of accumulation accelerates convergence. As in the original WL algorithm, our algorithm has two additional control parameters: the flatness criterion for the temperature histogram and the terminal value $f_\text{fin}$ of the modification factor. We consider the histogram to be flat when its smallest entry exceeds $80\%$ of the average histogram value~\cite{Wang2001}. All the results presented below are obtained using this flatness criterion. For $f_\text{fin}$, on the other hand, we systematically study its impact on the performance of subsequent equilibrium ST simulations in the next section. The value of $f_\text{fin}$ to achieve a given level of precision strongly depends on the model, the temperature range, and the system size. We thus need to carefully choose the terminal value so as to avoid a strongly biased temperature distribution for the equilibrium ST runs.

    \begin{figure}[t]
        \centering
        \includegraphics[width=\linewidth]{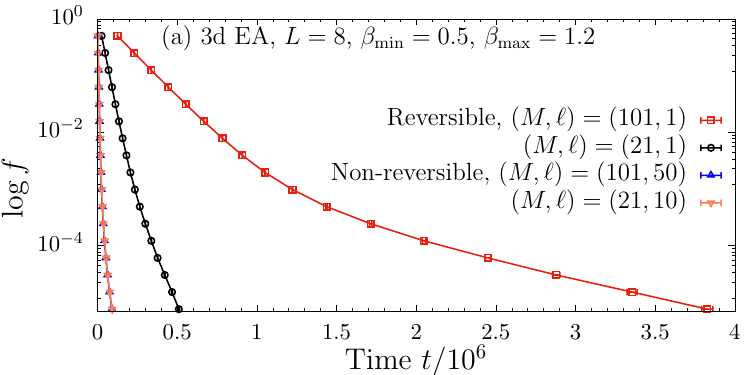}
        \includegraphics[width=\linewidth]{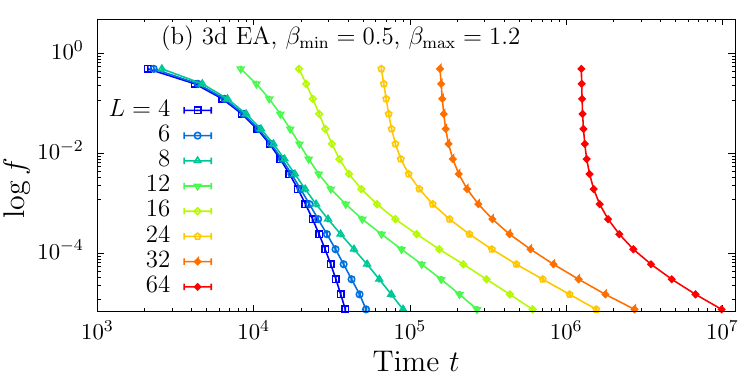}
        \caption{Time dependence of the modification factor $\log f$ during WLST free-energy estimation. (a) Comparison between the reversible and nonreversible WLST algorithms for $L=8$, shown for two choices of $(M, \ell)$. (b) Results for the nonreversible WLST algorithm at several system sizes, with $(M, \ell) = (101, 50)$. }
        \label{fig: log f vs time for many L}
    \end{figure}

    \section{Numerical tests on Edwards-Anderson spin glass model}
    \label{sec: numerical results on EASG model}

    We numerically test our algorithm on the three-dimensional Edwards-Anderson spin-glass model \cite{Edwards1975theory}. The Hamiltonian of the system is given by 
    \begin{align}
        H(\vec \sigma) = -\sum_{\langle i, j \rangle} J_{ij} \sigma_i \sigma_j,
    \end{align}
    where $\sigma_i \in \{\pm 1\}$ is a classical Ising spin at lattice site $i$ on a three-dimensional periodic cubic lattice with linear length $L$, and the summation runs over all nearest-neighbor pairs. The interaction constants $J_{ij}$ between spin $i$ and $j$ are independently drawn from a Gaussian distribution with zero mean and variance $J^2$. The model has a spin-glass transition at the critical temperature $\beta_c J = 1.05(1)$ \cite{Katzgraber2006universality}. For the local dynamics, the dynamical exponent $z$ is equal to $6.45(10)$~\cite{Katzgraber2005dynamical}. 
    
    We apply the WLST algorithm to the model with an inverse-temperature range $(\beta_\text{min}J, \beta_\text{max}J) = (0.5, 1.2)$, which covers the estimated critical temperature. The lowest temperature lies well into the spin-glass phase, as $\beta_\text{max} \approx 1.1\beta_c$. We restrict the inverse temperature of the system to one of $M = 101$ equally spaced values $\{\beta_i\}$, with $i$-th one set to
    \begin{align}
      \beta_i=\beta_\text{min} + i \times \frac{\beta_\text{max} - \beta_\text{min}}{M - 1}, \quad i=0,1,\ldots,M-1.       
    \end{align}
    Our WLST simulation starts with $\log f = 1$ and continues until $\log f$ reaches a predetermined terminal value $\log f_\text{fin}$. For each system size, disorder averages of $\{J_{ij}\}$ are taken over $512$ random realizations. 

    We benchmark the WLST algorithm against an on-the-fly thermodynamic integration (TI) algorithm proposed by Nguyen, Okamoto, and Derreumaux \cite{Nguyen2013simulated}, which we refer to as the NOD-TI algorithm hereafter. This algorithm follows a simple sequential procedure for estimating the free energy: The system is first quenched to the prescribed maximum inverse temperature $\beta_{M-1}$ and the energy samples are accumulated to estimate the equilibrium mean energy. We set $\hat \varphi(\beta_{M-1})=0$ and use $\hat \varphi(\beta_i)$ to denote an estimate for the free energy at $\beta_{M-1}$.
    An estimate at the neighboring inverse temperature $\beta_{M-2}$ is obtained as \cite{Park2007choosing}
    \begin{align}
        \hat \varphi(\beta_{M-2}) = (\beta_{M-2} - \beta_{M-1}) \hat E_{M-1},
    \end{align}
    with $\hat E_{M-1}$ an estimated mean energy at $\beta = \beta_{M-1}$. While accumulating energy samples, the algorithm continues attempts to visit $\beta_{M-2}$. However, as the mean-energy estimate $\hat E_{M-1}$ is not precise at the initial stage, virtually all the attempts are rejected. Only after the system is well equilibrated at $\beta_{M-1}$, the next inverse temperature $\beta_{M-2}$ becomes accessible and the algorithm starts to estimate the mean energy there. By repeating this process with gradually decreasing accessible inverse temperatures, we compute a free-energy estimate at $\beta_{i-1}$ as
    \begin{align}
        \hat \varphi(\beta_{i-1}) = \hat \varphi(\beta_{i}) + (\beta_{i-1} - \beta_{i}) \frac{\hat E_{i-1} + \hat E_{i}}{2}.
    \end{align}
    The algorithm terminates when it reaches $\beta_0$ and the free-energy estimate converges. We implement this algorithm using the same nonreversible dynamical rule in temperature space with $\ell = M/2$.
    
    The NOD-TI algorithm, despite its simplicity, works well on some models \cite{Nguyen2013simulated}. However, its performance is clearly controlled by the equilibration timescale at $\beta_{M-1}$. For systems that are hard to equilibrate at low temperatures, convergence can be extremely slow or may not be achieved within accessible simulation times. Another limitation of the original NOD-TI algorithm is that the precision of the free-energy estimate is not quantitatively controlled. Here, we modify the algorithm so that it halts when 
    \begin{align}
        \left|\frac{\hat \varphi_{k+1}(\beta_i) - \hat \varphi_k(\beta_i)}{\hat \varphi_{k+1}(\beta_i)}\right| < r_\text{tol},
        \label{eq: tolerance for NOD}
    \end{align}
    for all $i \in \{0, 1, \cdots, M-2\}$, where $\hat \varphi_{k}(\beta_i)$ is an instantaneous free-energy estimate at inverse temperature $\beta_i$ after the $k$-th trip through temperature space. We systematically change the tolerance $r_\text{tol}$ to control the precision of the final free-energy estimate.

    \begin{figure}[t]
        \centering
        \includegraphics[width=\linewidth]{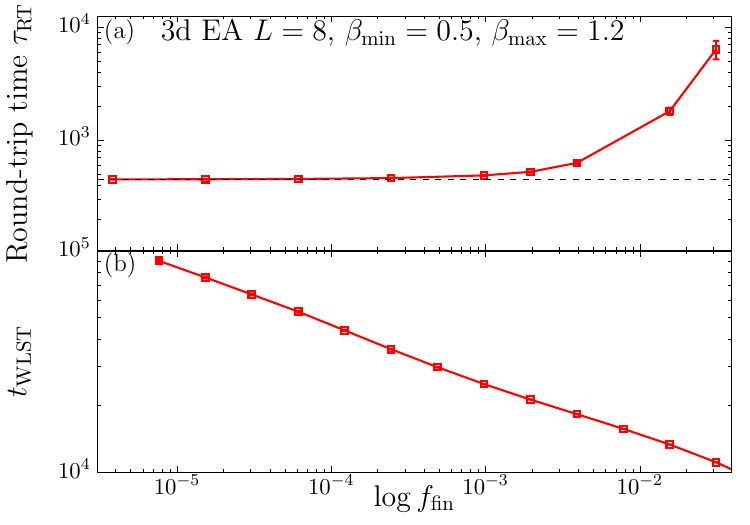}
        \caption{(a) Round-trip time $\tau_\text{RT}$ measured in equilibrium ST simulations and (b) WLST simulation time $t_\text{WLST}$ required to reach the terminal value $\log f_\text{fin}$, both as a function of $\log f_{\mathrm{fin}}$. From these plots, $\tau_\text{RT}$ is given as a function of $t_\text{WLST}$, yielding the convergence time $\tau_c^{\mathrm{WLST}}$ with \eq{eq: tau_c for WLST}. The system size is $L = 8$ and the innverse-temperature range is $(\beta_\text{min}J, \beta_\text{max}J) = (0.5, 1.2)$.}
        \label{fig: round-trip time and t_wlst vs log f}
    \end{figure}

    We first compare our nonreversible WLST algorithm with its conventional reversible counterpart, for which $\ell = 1$. \fig{fig: log f vs time for many L}~(a) shows the time dependencies of the modification parameter $\log f$ for the two algorithms at the system size $L = 8$. The timescale of the decay in $\log f$ for the nonreversible algorithm is almost independent of $M$, whereas that for the reversible one increases strongly with $M$. This timescale for the nonreversible algorithm, on the other hand, has a clear dependence on the system size. In \fig{fig: log f vs time for many L}~(b), we show the time dependence of the control parameter $\log f$ for the nonreversible WLST algorithm with $(M,\ell)=(101,50)$ at several system sizes. For small sizes $L \lesssim 8$, $\log f$ decreases nearly exponentially, resulting in the rapid convergence. With increasing $L$, the decay becomes slower and crosses over to an asymptotic power-law behavior, approximately $\log f\sim t^{-1.6}$ for $L \gtrsim 12$. 

    In the WLST algorithm, the final modification factor $\log f_\text{fin}$ controls the precision of the free-energy estimate; a smaller value for $\log f_\text{fin}$ yields a more precise estimate. If the bias function $g(\beta)$ strongly deviates from $\varphi(\beta)$, the temperature distribution of equilibrium ST simulations is far from uniform, and the dynamics in the temperature space should be suboptimal. To quantify the effects of nonzero $\log f_\text{fin}$, we run equilibrium ST simulations using $g(\beta)$ set to the free-energy estimate and measure the mean round-trip time $\tau_\text{RT}$, that is, the mean number of Monte Carlo sweeps per spin needed for the temperature variable to travel between the lowest and highest $\beta$s. The round-trip time $\tau_\text{RT}$ decays and converges when $\log f_\text{min}$ decreases (see \fig{fig: round-trip time and t_wlst vs log f}~(a)), showing that setting $g(\beta)$ to $\varphi(\beta)$ is indeed optimal.
    
    To quantitatively compare the WLST and the NOD-TI algorithms, we compute the round-trip time as a function of the computational effort needed by each algorithm, measured by the number of sweeps per spin, $t_\text{WLST}$ and $t_\text{NOD}$, respectively. Recall that the round-trip time characterizes a timescale of equilibrium ST simulations with fixed $g(\beta)$, whereas $t_\text{WLST}$ and $t_\text{NOD}$ refer to the number of sweeps while modifying $g(\beta)$. To this end, we first calculate $t_\text{WLST}$ to achieve a given $\log f_\text{fin}$ (\fig{fig: round-trip time and t_wlst vs log f}~(b)). This allows us to represent the round-trip time $\tau_\text{RT}$ as a function of $t_\text{WLST}$. We then define the operational convergence time $\tau_c^\text{WLST}$ as $t_\text{WLST}$ that satisfies
    \begin{align}
        \frac{\tau_\text{RT}(t_\text{WLST} = \tau^\text{WLST}_c) - \tau^\infty_\text{RT}}{\tau^\infty_\text{RT}} = 10^{-1},
        \label{eq: tau_c for WLST}
    \end{align}
    with $\tau^\infty_\text{RT}$ the limiting round-trip time at $\log f_\text{fin} \to 0$. The convergence time $\tau^\text{NOD}_c$ for the NOD-TI algorithm is defined similarly, through measuring $t_\text{NOD}$ to reach a given level of precision specified by $r_\text{tol}$ in \eq{eq: tolerance for NOD} and $\tau_\text{RT}$ as a function of $r_\text{tol}$. In practice, we set $\tau^\infty_\text{RT}$ to the round-trip time obtained by using the free-energy estimates of the WLST algorithm at $\log f = 2^{-19}$, $2^{-20}$, and $2^{-21}$, for $L \leq 16$, $L = 18$, and $L \geq 20$, respectively. We have confirmed that for each system size, $\tau_\text{RT}$ does not strongly depend on $\log f_\text{fin}$ when reaching these values. 

    \begin{figure}[t]
        \includegraphics[width=\linewidth]{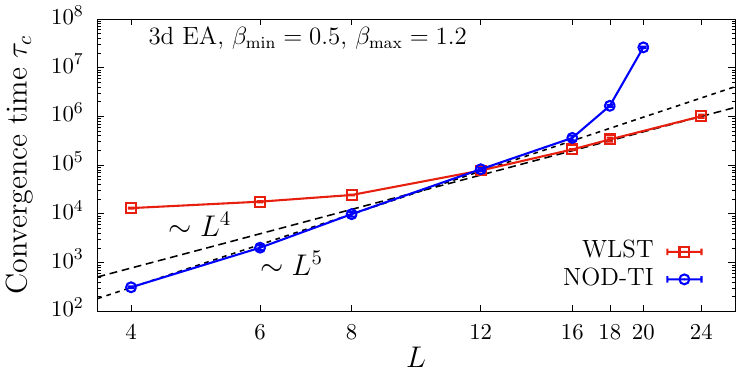}
        \caption{Convergence time $\tau_c$ as a function of the system size $L$ for the WLST and NOD-TI algorithms. 
        The dashed and dotted lines indicate the power-law scalings $L^4$ and $L^5$, respectively. 
        }
        \label{fig: convergence time for ising sg}
    \end{figure}

    We show the convergence time $\tau_c$ of the two algorithms as functions of $L$ in \fig{fig: convergence time for ising sg}. For small system sizes $L \lesssim 12$, where equilibration at the largest $\beta$ is not too difficult, the NOD-TI algorithm converges faster than the WLST algorithm. When increasing $L$, the values for $\tau_c$ of both algorithms grow, but that of the NOD-TI algorithm exhibits a stronger dependence on $L$: For $L\gtrsim 12$, $\tau_c$ for the WLST algorithm follows an approximate power law, $\tau_c^{\mathrm{WLST}} \sim L^4$, while $\tau_c^{\mathrm{NOD}} \sim L^5$ for the NOD-TI, and crosses over to an almost exponential growth for $L \gtrsim 16$. This drastic growth of $\tau_c^{\mathrm{NOD}}$ originates from the fact that the system is first quenched to the largest inverse temperature $\beta_{M-1}$ and stays there until it reaches or approaches close enough to equilibrium. The convergence time of the NOD-TI algorithm is thus inherently bounded from below by the equilibration timescale at $\beta_{M-1}$. The WLST algorithm, on the other hand, can travel through the entire range of $\beta$ even at an early stage of the free-energy estimation, accelerating equilibration at $\beta_{M-1}$ and leading to a weaker dependence on $L$. 

    \begin{figure}[t]
        \centering
        \includegraphics[width=\linewidth]{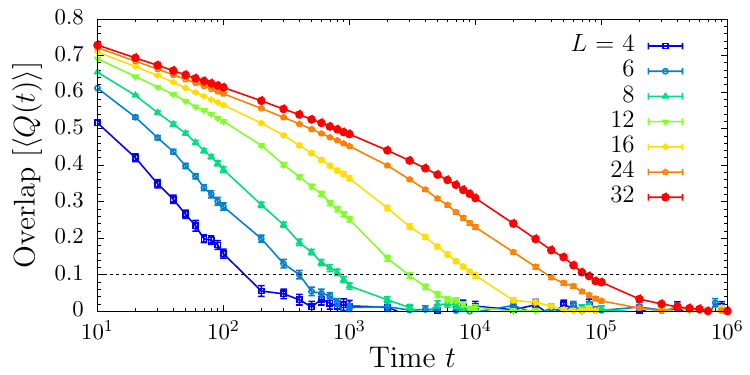}
        \caption{Time-dependent overlap $[\average{Q(t)}]$ measured along ST trajectories with fluctuating inverse temperature $\beta$, as a function of time $t$ for several system sizes $L$. Each trajectory is initialized with an equilibrium configuration at $\beta J = 1.2$. The horizontal dashed line indicates the threshold $[\average{Q(\tau_s)}] = 10^{-1}$ used to define the relaxation time $\tau_s$. }
        \label{fig: overlap}
    \end{figure}

    \begin{figure}[t]
        \centering
        \includegraphics[width=\linewidth]{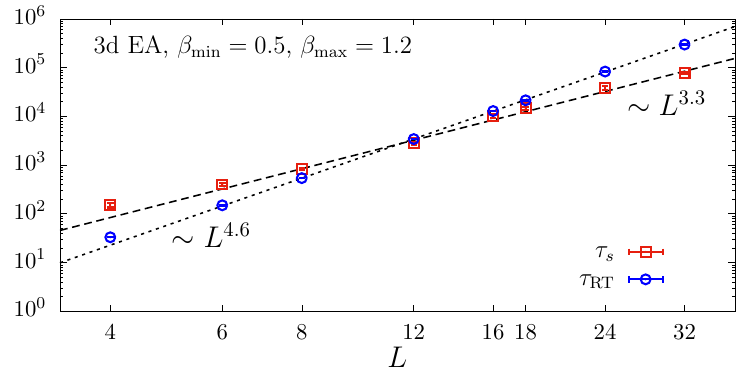}
        \caption{Relaxation time $\tau_s$ and round-trip time $\tau_\text{RT}$ as functions of $L$. Power-law fits defined by $\tau_s\sim L^{z_s}$ and $\tau_{\mathrm{RT}}\sim L^{z_{\mathrm{RT}}}$ yield the effective exponents $z_s \approx 3.3$ for $\tau_s$ and $z_{\mathrm{RT}} \approx 4.6$ for $\tau_\text{RT}$.}
        \label{fig: other time scales}
    \end{figure}

    We next turn to a discussion on the performance of the nonreversible ST simulations for the spin-glass model. In addition to the round-trip time $\tau_\text{RT}$ introduced above, we define the relaxation time $\tau_s$ from the time-dependent overlap
    \begin{align}
        [\average{Q(t)}] = \left[\average{\frac1N \sum_i \sigma_i(0)\sigma_i(t)} \right].
    \end{align}
    Here, $\average{\cdots}$ denotes an average over equilibrium initial spin configurations at $\beta_{M-1}$ and trajectories for a fixed realization of the interactions, and $[\cdots]$ represents a disorder average over $\{J_{ij}\}$. One unit time consists of one Monte Carlo sweep per spin followed by $\ell$ attempted temperature updates. We take an equilibrium spin configuration at the largest inverse temperature $\beta_{M-1}$ as the initial configuration $\vec \sigma$ and set $\beta = \beta_{M-1}$ at time $t = 0$, and measure the overlap at each subsequent time with fluctuating $\beta$. 
    
    In \fig{fig: overlap}, we show the overlap $[\average{Q(t)}]$ for $L = 4, 6, \cdots, 32$. The decay becomes slower as $L$ increases, indicating a growing relaxation timescale. We define the relaxation time $\tau_s$ as the time where $[\average{Q(t = \tau_s)}] = 10^{-1}$ and plot it as a function of $L$ in \fig{fig: other time scales}, together with the round-trip time $\tau_\text{RT}$. Both timescales exhibit clear power-law growths $\tau_X \sim L^{z_X}$ ($X \in \{s, \text{RT}\}$), with exponents $z_s \approx 3.3$ and $z_\text{RT} \approx 4.6$, respectively. The total computational cost of the nonreversible ST, including the preliminary stage for estimating the free energy, is thus $O(L^{z_s}) + O(L^{z_\text{RT}}) + O(L^4) = O(L^{z_\text{RT}})$ for our setup. Whereas these exponents should depend on the parameters and do not originate only from the nature of the spin-glass phase of the model, these allow us to directly compare the computational efficiency of the nonreversible ST and that of the local dynamics: The relaxation time of $[\average{Q(t)}]$ for the local Metropolis dynamics at $\beta = \beta_c$ grows algebraically as $L$ increases, with exponent $z_\text{local} \approx 6.45$ \cite{Katzgraber2005dynamical}. The three exponents for the nonreversible ST, all of which are significantly smaller than $z_\text{local}$, thus imply a large computational advantage of the algorithm for large system sizes in the spin glass phase.

    \section{Conclusion}
    \label{sec: conclusion}
    In this paper, we have presented an efficient ST-based algorithm for free-energy estimation that integrates the Wang-Landau-type feedback and lifted nonreversible dynamics in temperature space. The lifted dynamics naturally accommodates multiple attempts for updating temperature per tempering step, generating persistent and directed dynamics through temperature space. During these processes, the temperature histogram and the free-energy estimate are modified after every attempt, which also accelerates convergence of the Wang--Landau-type feedback. 
    We have benchmarked the proposed method by applying it to the three-dimensional Edwards-Anderson spin-glass model. Our results show that, thanks to the nonreversible dynamical rule and the Wang--Landau-type feedback, the algorithm rapidly covers the whole prescribed temperature range while estimating the free energy, and that the estimate converges substantially faster than that of the NOD-TI algorithm \cite{Nguyen2013simulated}, which initially quenches the system to the lowest temperature. We successfully estimated the free energy over a broad temperature range extending well into the spin-glass phase for system sizes up to $L\simeq 64$.  
    A further practical advantage is that the free energy is estimated within a single adaptive simulation stage, without requiring multiple stages of simulations and subsequent statistical analyses using, e.g., the reweighting technique \cite{Ferrenberg1988new,Ferrenberg1989optimized,Wang2015measuring}, allowing a seamless switching to equilibrium ST production runs. 
    These results show that the proposed method significantly reduces one of the main practical barriers to applying ST to large-scale complex systems. 

    Our lifted nonreversible update rule may seem more complicated than the conventional reversible temperature updates, but its implementation is actually simple and straightforward, with only single one additional parameter, that is, the lifting variable. Because the moving direction in temperature space is fully determined by the lifting variable, each nonreversible update requires fewer pseudorandom numbers per step than the reversible one. Therefore, significant speedup will be attained with only minor modifications to an ST code. While PT has traditionally been regarded as particularly well-suited to parallel computation, ST, in principle, always allows for the parallel implementation of multiple independent simulations. However, in practice, it has remained difficult to take advantage of this feature because precisely estimating the free energy is inherently challenging. The proposed method largely reduces this difficulty, making communication-free, massive parallelization of ST practical. In particular, for a given system, multiple independent production runs can proceed in parallel using the same estimated bias function, without requiring communication or synchronization between them. This feature makes ST highly suitable for large-scale equilibrium state sampling on massively parallel computers. 

    Our nonreversible ST algorithm is competitive even when compared with an optimized parallel tempering run: Our preliminary results on the same spin glass model for $(\beta_\text{min}J, \beta_\text{max}J) = (0.5, 5)$, which Ref.~\cite{Yucesoy2013correlations} uses for their parallel tempering runs, suggest that $\tau_\text{RT}$ grows significantly slower for the nonreversible ST for $L$ up to $10$ than that for their optimized parallel tempering (not shown). We argue that this weaker dependence on $L$ is achieved by the maximally nonreversible nature of the algorithm and allows us to study the spin-glass transition more systematically than before with large system sizes. The advantages of our algorithm are, however, not expected to extend universally to other complex systems. Systems with a first-order transition are typical examples, in which ST cannot overcome a large free-energy barrier and has exponentially slow relaxation. Other complex systems without such strong first-order bottlenecks, including some structural-glass and polymer models, could benefit from the nonreversible nature of the algorithm. Understanding what physical properties of these systems control the performance of optimized tempering algorithms is an important direction for future work.

    \begin{acknowledgments}
        We thank A.~C.~Maggs, S.~Takabe and L.~Berthier for useful discussions. This work is supported by JSPS KAKENHI (Grant No.~22K13968, 23H01095, 26K14992, and JST Grant Number JPMJPF2221). The computations in this work were partially performed using the facilities of the Supercomputer Center, the Institute for Solid State Physics, the University of Tokyo (ISSPkyodo-SC-2024-Cb-0041, 2025-Ca-0081).
    \end{acknowledgments}
    \bibliography{refs}

\begin{thebibliography}{57}%
\makeatletter
\providecommand \@ifxundefined [1]{%
 \@ifx{#1\undefined}
}%
\providecommand \@ifnum [1]{%
 \ifnum #1\expandafter \@firstoftwo
 \else \expandafter \@secondoftwo
 \fi
}%
\providecommand \@ifx [1]{%
 \ifx #1\expandafter \@firstoftwo
 \else \expandafter \@secondoftwo
 \fi
}%
\providecommand \natexlab [1]{#1}%
\providecommand \enquote  [1]{``#1''}%
\providecommand \bibnamefont  [1]{#1}%
\providecommand \bibfnamefont [1]{#1}%
\providecommand \citenamefont [1]{#1}%
\providecommand \href@noop [0]{\@secondoftwo}%
\providecommand \href [0]{\begingroup \@sanitize@url \@href}%
\providecommand \@href[1]{\@@startlink{#1}\@@href}%
\providecommand \@@href[1]{\endgroup#1\@@endlink}%
\providecommand \@sanitize@url [0]{\catcode `\\12\catcode `\$12\catcode
  `\&12\catcode `\#12\catcode `\^12\catcode `\_12\catcode `\%12\relax}%
\providecommand \@@startlink[1]{}%
\providecommand \@@endlink[0]{}%
\providecommand \url  [0]{\begingroup\@sanitize@url \@url }%
\providecommand \@url [1]{\endgroup\@href {#1}{\urlprefix }}%
\providecommand \urlprefix  [0]{URL }%
\providecommand \Eprint [0]{\href }%
\providecommand \doibase [0]{https://doi.org/}%
\providecommand \selectlanguage [0]{\@gobble}%
\providecommand \bibinfo  [0]{\@secondoftwo}%
\providecommand \bibfield  [0]{\@secondoftwo}%
\providecommand \translation [1]{[#1]}%
\providecommand \BibitemOpen [0]{}%
\providecommand \bibitemStop [0]{}%
\providecommand \bibitemNoStop [0]{.\EOS\space}%
\providecommand \EOS [0]{\spacefactor3000\relax}%
\providecommand \BibitemShut  [1]{\csname bibitem#1\endcsname}%
\let\auto@bib@innerbib\@empty
\bibitem [{\citenamefont {Swendsen}\ and\ \citenamefont
  {Wang}(1987)}]{Swendsen1987nonuniversal}%
  \BibitemOpen
  \bibfield  {author} {\bibinfo {author} {\bibfnamefont {R.~H.}\ \bibnamefont
  {Swendsen}}\ and\ \bibinfo {author} {\bibfnamefont {J.-S.}\ \bibnamefont
  {Wang}},\ }\bibfield  {title} {\bibinfo {title} {{Nonuniversal Critical
  Dynamics in Monte Carlo Simulations}},\ }\href
  {https://doi.org/10.1103/PhysRevLett.58.86} {\bibfield  {journal} {\bibinfo
  {journal} {Physical Review Letters}\ }\textbf {\bibinfo {volume} {58}},\
  \bibinfo {pages} {86} (\bibinfo {year} {1987})}\BibitemShut {NoStop}%
\bibitem [{\citenamefont {Wolff}(1989)}]{Wolff1989collective}%
  \BibitemOpen
  \bibfield  {author} {\bibinfo {author} {\bibfnamefont {U.}~\bibnamefont
  {Wolff}},\ }\bibfield  {title} {\bibinfo {title} {{Collective Monte Carlo
  Updating for Spin Systems}},\ }\href
  {https://doi.org/10.1103/PhysRevLett.62.361} {\bibfield  {journal} {\bibinfo
  {journal} {Physical Review Letters}\ }\textbf {\bibinfo {volume} {62}},\
  \bibinfo {pages} {361} (\bibinfo {year} {1989})}\BibitemShut {NoStop}%
\bibitem [{\citenamefont {Dress}\ and\ \citenamefont
  {Krauth}(1995)}]{Dress1996cluster}%
  \BibitemOpen
  \bibfield  {author} {\bibinfo {author} {\bibfnamefont {C.}~\bibnamefont
  {Dress}}\ and\ \bibinfo {author} {\bibfnamefont {W.}~\bibnamefont {Krauth}},\
  }\bibfield  {title} {\bibinfo {title} {{Cluster algorithm for hard spheres
  and related systems}},\ }\href {https://doi.org/10.1088/0305-4470/28/23/001}
  {\bibfield  {journal} {\bibinfo  {journal} {Journal of Physics A:
  Mathematical and General}\ }\textbf {\bibinfo {volume} {28}},\ \bibinfo
  {pages} {L597} (\bibinfo {year} {1995})}\BibitemShut {NoStop}%
\bibitem [{\citenamefont {Liu}\ and\ \citenamefont
  {Luijten}(2004)}]{Liu2004rejection}%
  \BibitemOpen
  \bibfield  {author} {\bibinfo {author} {\bibfnamefont {J.}~\bibnamefont
  {Liu}}\ and\ \bibinfo {author} {\bibfnamefont {E.}~\bibnamefont {Luijten}},\
  }\bibfield  {title} {\bibinfo {title} {{Rejection-Free Geometric Cluster
  Algorithm for Complex Fluids}},\ }\href
  {https://doi.org/10.1103/PhysRevLett.92.035504} {\bibfield  {journal}
  {\bibinfo  {journal} {Physical Review Letters}\ }\textbf {\bibinfo {volume}
  {92}},\ \bibinfo {pages} {035504} (\bibinfo {year} {2004})}\BibitemShut
  {NoStop}%
\bibitem [{\citenamefont {Iba}(2001)}]{Iba_2001}%
  \BibitemOpen
  \bibfield  {author} {\bibinfo {author} {\bibfnamefont {Y.}~\bibnamefont
  {Iba}},\ }\bibfield  {title} {\bibinfo {title} {Extended ensemble monte
  carlo},\ }\href {https://doi.org/10.1142/S0129183101001912} {\bibfield
  {journal} {\bibinfo  {journal} {International Journal of Modern Physics C}\
  }\textbf {\bibinfo {volume} {12}},\ \bibinfo {pages} {623} (\bibinfo {year}
  {2001})},\ \Eprint
  {https://arxiv.org/abs/https://doi.org/10.1142/S0129183101001912}
  {https://doi.org/10.1142/S0129183101001912} \BibitemShut {NoStop}%
\bibitem [{\citenamefont {Marinari}\ and\ \citenamefont
  {Parisi}(1992)}]{Marinari1992}%
  \BibitemOpen
  \bibfield  {author} {\bibinfo {author} {\bibfnamefont {E.}~\bibnamefont
  {Marinari}}\ and\ \bibinfo {author} {\bibfnamefont {G.}~\bibnamefont
  {Parisi}},\ }\bibfield  {title} {\bibinfo {title} {{Simulated Tempering: A
  New Monte Carlo Scheme}},\ }\href
  {https://doi.org/10.1209/0295-5075/19/6/002} {\bibfield  {journal} {\bibinfo
  {journal} {Europhysics Letters (EPL)}\ }\textbf {\bibinfo {volume} {19}},\
  \bibinfo {pages} {451} (\bibinfo {year} {1992})}\BibitemShut {NoStop}%
\bibitem [{\citenamefont {Lyubartsev}\ \emph {et~al.}(1992)\citenamefont
  {Lyubartsev}, \citenamefont {Martsinovski}, \citenamefont {Shevkunov},\ and\
  \citenamefont {Vorontsov‐Velyaminov}}]{Lyubartsev1992}%
  \BibitemOpen
  \bibfield  {author} {\bibinfo {author} {\bibfnamefont {A.~P.}\ \bibnamefont
  {Lyubartsev}}, \bibinfo {author} {\bibfnamefont {A.~A.}\ \bibnamefont
  {Martsinovski}}, \bibinfo {author} {\bibfnamefont {S.~V.}\ \bibnamefont
  {Shevkunov}},\ and\ \bibinfo {author} {\bibfnamefont {P.~N.}\ \bibnamefont
  {Vorontsov‐Velyaminov}},\ }\bibfield  {title} {\bibinfo {title} {New
  approach to monte carlo calculation of the free energy: Method of expanded
  ensembles},\ }\href {https://doi.org/10.1063/1.462133} {\bibfield  {journal}
  {\bibinfo  {journal} {The Journal of Chemical Physics}\ }\textbf {\bibinfo
  {volume} {96}},\ \bibinfo {pages} {1776} (\bibinfo {year}
  {1992})}\BibitemShut {NoStop}%
\bibitem [{\citenamefont {Mykland}\ \emph {et~al.}(1995)\citenamefont
  {Mykland}, \citenamefont {Tierney},\ and\ \citenamefont
  {Yu}}]{Mykland1995regeneration}%
  \BibitemOpen
  \bibfield  {author} {\bibinfo {author} {\bibfnamefont {P.}~\bibnamefont
  {Mykland}}, \bibinfo {author} {\bibfnamefont {L.}~\bibnamefont {Tierney}},\
  and\ \bibinfo {author} {\bibfnamefont {B.}~\bibnamefont {Yu}},\ }\bibfield
  {title} {\bibinfo {title} {{Regeneration in Markov Chain Samplers}},\ }\href
  {https://doi.org/10.2307/2291148} {\bibfield  {journal} {\bibinfo  {journal}
  {Journal of the American Statistical Association}\ }\textbf {\bibinfo
  {volume} {90}},\ \bibinfo {pages} {233} (\bibinfo {year} {1995})}\BibitemShut
  {NoStop}%
\bibitem [{\citenamefont {Biron-Lattes}\ \emph {et~al.}(2025)\citenamefont
  {Biron-Lattes}, \citenamefont {Campbell},\ and\ \citenamefont
  {Bouchard-C{\^{o}}t{\'{e}}}}]{Biron-Lattes2025automatic}%
  \BibitemOpen
  \bibfield  {author} {\bibinfo {author} {\bibfnamefont {M.}~\bibnamefont
  {Biron-Lattes}}, \bibinfo {author} {\bibfnamefont {T.}~\bibnamefont
  {Campbell}},\ and\ \bibinfo {author} {\bibfnamefont {A.}~\bibnamefont
  {Bouchard-C{\^{o}}t{\'{e}}}},\ }\bibfield  {title} {\bibinfo {title}
  {{Automatic Regenerative Simulation via Non-Reversible Simulated
  Tempering}},\ }\href {https://doi.org/10.1080/01621459.2024.2335587}
  {\bibfield  {journal} {\bibinfo  {journal} {Journal of the American
  Statistical Association}\ }\textbf {\bibinfo {volume} {120}},\ \bibinfo
  {pages} {318} (\bibinfo {year} {2025})}\BibitemShut {NoStop}%
\bibitem [{\citenamefont {Geyer}(1991)}]{Geyer1991}%
  \BibitemOpen
  \bibfield  {author} {\bibinfo {author} {\bibfnamefont {C.~J.}\ \bibnamefont
  {Geyer}},\ }\bibfield  {title} {\bibinfo {title} {Markov chain monte carlo
  maximum likelihood},\ }in\ \href@noop {} {\emph {\bibinfo {booktitle}
  {Computing Science and Statistics: Proceedings of the 23rd Symposium on the
  Interface}}}\ (\bibinfo {year} {1991})\ pp.\ \bibinfo {pages}
  {156--163}\BibitemShut {NoStop}%
\bibitem [{\citenamefont {Hukushima}\ and\ \citenamefont
  {Nemoto}(1996)}]{Hukushima1996}%
  \BibitemOpen
  \bibfield  {author} {\bibinfo {author} {\bibfnamefont {K.}~\bibnamefont
  {Hukushima}}\ and\ \bibinfo {author} {\bibfnamefont {K.}~\bibnamefont
  {Nemoto}},\ }\bibfield  {title} {\bibinfo {title} {{Exchange Monte Carlo
  Method and Application to Spin Glass Simulations}},\ }\href
  {https://doi.org/10.1143/JPSJ.65.1604} {\bibfield  {journal} {\bibinfo
  {journal} {Journal of the Physical Society of Japan}\ }\textbf {\bibinfo
  {volume} {65}},\ \bibinfo {pages} {1604} (\bibinfo {year}
  {1996})}\BibitemShut {NoStop}%
\bibitem [{\citenamefont {Marinari}(1998)}]{Marinari1998}%
  \BibitemOpen
  \bibfield  {author} {\bibinfo {author} {\bibfnamefont {E.}~\bibnamefont
  {Marinari}},\ }\bibfield  {title} {\bibinfo {title} {Optimized monte carlo
  methods},\ }in\ \href@noop {} {\emph {\bibinfo {booktitle} {Advances in
  Computer Simulation}}},\ \bibinfo {series} {Lecture Notes in Physics}, Vol.\
  \bibinfo {volume} {501},\ \bibinfo {editor} {edited by\ \bibinfo {editor}
  {\bibfnamefont {J.}~\bibnamefont {Kert{\'e}sz}}\ and\ \bibinfo {editor}
  {\bibfnamefont {I.}~\bibnamefont {Kondor}}}\ (\bibinfo  {publisher}
  {Springer},\ \bibinfo {year} {1998})\ pp.\ \bibinfo {pages}
  {50--81}\BibitemShut {NoStop}%
\bibitem [{\citenamefont {Mitsutake}\ and\ \citenamefont
  {Okamoto}(2000)}]{Mitsutake2000replica}%
  \BibitemOpen
  \bibfield  {author} {\bibinfo {author} {\bibfnamefont {A.}~\bibnamefont
  {Mitsutake}}\ and\ \bibinfo {author} {\bibfnamefont {Y.}~\bibnamefont
  {Okamoto}},\ }\bibfield  {title} {\bibinfo {title} {{Replica-exchange
  simulated tempering method for simulations of frustrated systems}},\ }\href
  {https://doi.org/10.1016/S0009-2614(00)01262-8} {\bibfield  {journal}
  {\bibinfo  {journal} {Chemical Physics Letters}\ }\textbf {\bibinfo {volume}
  {332}},\ \bibinfo {pages} {131} (\bibinfo {year} {2000})}\BibitemShut
  {NoStop}%
\bibitem [{\citenamefont {Mitsutake}\ and\ \citenamefont
  {Okamoto}(2004)}]{Mitsutake2004replica}%
  \BibitemOpen
  \bibfield  {author} {\bibinfo {author} {\bibfnamefont {A.}~\bibnamefont
  {Mitsutake}}\ and\ \bibinfo {author} {\bibfnamefont {Y.}~\bibnamefont
  {Okamoto}},\ }\bibfield  {title} {\bibinfo {title} {{Replica-exchange
  extensions of simulated tempering method}},\ }\href
  {https://doi.org/10.1063/1.1766015} {\bibfield  {journal} {\bibinfo
  {journal} {Journal of Chemical Physics}\ }\textbf {\bibinfo {volume} {121}},\
  \bibinfo {pages} {2491} (\bibinfo {year} {2004})}\BibitemShut {NoStop}%
\bibitem [{\citenamefont {Park}\ and\ \citenamefont
  {Pande}(2007)}]{Park2007choosing}%
  \BibitemOpen
  \bibfield  {author} {\bibinfo {author} {\bibfnamefont {S.}~\bibnamefont
  {Park}}\ and\ \bibinfo {author} {\bibfnamefont {V.~S.}\ \bibnamefont
  {Pande}},\ }\bibfield  {title} {\bibinfo {title} {{Choosing weights for
  simulated tempering}},\ }\href {https://doi.org/10.1103/PhysRevE.76.016703}
  {\bibfield  {journal} {\bibinfo  {journal} {Physical Review E}\ }\textbf
  {\bibinfo {volume} {76}},\ \bibinfo {pages} {016703} (\bibinfo {year}
  {2007})}\BibitemShut {NoStop}%
\bibitem [{\citenamefont {Nguyen}\ \emph {et~al.}(2013)\citenamefont {Nguyen},
  \citenamefont {Okamoto},\ and\ \citenamefont
  {Derreumaux}}]{Nguyen2013simulated}%
  \BibitemOpen
  \bibfield  {author} {\bibinfo {author} {\bibfnamefont {P.~H.}\ \bibnamefont
  {Nguyen}}, \bibinfo {author} {\bibfnamefont {Y.}~\bibnamefont {Okamoto}},\
  and\ \bibinfo {author} {\bibfnamefont {P.}~\bibnamefont {Derreumaux}},\
  }\bibfield  {title} {\bibinfo {title} {{Communication: Simulated tempering
  with fast on-the-fly weight determination}},\ }\href
  {https://doi.org/10.1063/1.4792046} {\bibfield  {journal} {\bibinfo
  {journal} {The Journal of Chemical Physics}\ }\textbf {\bibinfo {volume}
  {138}},\ \bibinfo {pages} {1} (\bibinfo {year} {2013})}\BibitemShut {NoStop}%
\bibitem [{\citenamefont {Sakai}\ and\ \citenamefont
  {Hukushima}(2016{\natexlab{a}})}]{Sakai2016learning}%
  \BibitemOpen
  \bibfield  {author} {\bibinfo {author} {\bibfnamefont {Y.}~\bibnamefont
  {Sakai}}\ and\ \bibinfo {author} {\bibfnamefont {K.}~\bibnamefont
  {Hukushima}},\ }\bibfield  {title} {\bibinfo {title} {{Irreversible simulated
  tempering algorithm with skew detailed balance conditions: a learning method
  of weight factors in simulated tempering}},\ }\href
  {https://doi.org/10.1088/1742-6596/750/1/012013} {\bibfield  {journal}
  {\bibinfo  {journal} {Journal of Physics: Conference Series}\ }\textbf
  {\bibinfo {volume} {750}},\ \bibinfo {pages} {012013} (\bibinfo {year}
  {2016}{\natexlab{a}})}\BibitemShut {NoStop}%
\bibitem [{\citenamefont {Ferrenberg}\ and\ \citenamefont
  {Swendsen}(1989)}]{Ferrenberg1989optimized}%
  \BibitemOpen
  \bibfield  {author} {\bibinfo {author} {\bibfnamefont {A.~M.}\ \bibnamefont
  {Ferrenberg}}\ and\ \bibinfo {author} {\bibfnamefont {R.~H.}\ \bibnamefont
  {Swendsen}},\ }\bibfield  {title} {\bibinfo {title} {{Optimized Monte Carlo
  data analysis}},\ }\href {https://doi.org/10.1103/PhysRevLett.63.1195}
  {\bibfield  {journal} {\bibinfo  {journal} {Physical Review Letters}\
  }\textbf {\bibinfo {volume} {63}},\ \bibinfo {pages} {1195} (\bibinfo {year}
  {1989})}\BibitemShut {NoStop}%
\bibitem [{\citenamefont {M{\"{u}}nger}\ and\ \citenamefont
  {Novotny}(1991)}]{Munger1991reweighting}%
  \BibitemOpen
  \bibfield  {author} {\bibinfo {author} {\bibfnamefont {E.~P.}\ \bibnamefont
  {M{\"{u}}nger}}\ and\ \bibinfo {author} {\bibfnamefont {M.~A.}\ \bibnamefont
  {Novotny}},\ }\bibfield  {title} {\bibinfo {title} {{Reweighting in Monte
  Carlo and Monte Carlo renormalization-group studies}},\ }\href
  {https://link.aps.org/doi/10.1103/PhysRevB.43.5773} {\bibfield  {journal}
  {\bibinfo  {journal} {Physical Review B}\ }\textbf {\bibinfo {volume} {43}},\
  \bibinfo {pages} {5773} (\bibinfo {year} {1991})}\BibitemShut {NoStop}%
\bibitem [{\citenamefont {Wang}\ and\ \citenamefont
  {Landau}(2001{\natexlab{a}})}]{Wang2001}%
  \BibitemOpen
  \bibfield  {author} {\bibinfo {author} {\bibfnamefont {F.}~\bibnamefont
  {Wang}}\ and\ \bibinfo {author} {\bibfnamefont {D.~P.}\ \bibnamefont
  {Landau}},\ }\bibfield  {title} {\bibinfo {title} {{Efficient, Multiple-Range
  Random Walk Algorithm to Calculate the Density of States}},\ }\href
  {https://doi.org/10.1103/PhysRevLett.86.2050} {\bibfield  {journal} {\bibinfo
   {journal} {Physical Review Letters}\ }\textbf {\bibinfo {volume} {86}},\
  \bibinfo {pages} {2050} (\bibinfo {year} {2001}{\natexlab{a}})}\BibitemShut
  {NoStop}%
\bibitem [{\citenamefont {Wang}\ and\ \citenamefont
  {Landau}(2001{\natexlab{b}})}]{WangLandau2001}%
  \BibitemOpen
  \bibfield  {author} {\bibinfo {author} {\bibfnamefont {F.}~\bibnamefont
  {Wang}}\ and\ \bibinfo {author} {\bibfnamefont {D.~P.}\ \bibnamefont
  {Landau}},\ }\bibfield  {title} {\bibinfo {title} {Determining the density of
  states for classical statistical models: A random walk algorithm to produce a
  flat histogram},\ }\href {https://doi.org/10.1103/PhysRevE.64.056101}
  {\bibfield  {journal} {\bibinfo  {journal} {Phys. Rev. E}\ }\textbf {\bibinfo
  {volume} {64}},\ \bibinfo {pages} {056101} (\bibinfo {year}
  {2001}{\natexlab{b}})}\BibitemShut {NoStop}%
\bibitem [{\citenamefont {Zhang}\ and\ \citenamefont
  {Ma}(2007)}]{Zhang2007simulation}%
  \BibitemOpen
  \bibfield  {author} {\bibinfo {author} {\bibfnamefont {C.}~\bibnamefont
  {Zhang}}\ and\ \bibinfo {author} {\bibfnamefont {J.}~\bibnamefont {Ma}},\
  }\bibfield  {title} {\bibinfo {title} {{Simulation via direct computation of
  partition functions}},\ }\href {https://doi.org/10.1103/PhysRevE.76.036708}
  {\bibfield  {journal} {\bibinfo  {journal} {Physical Review E}\ }\textbf
  {\bibinfo {volume} {76}},\ \bibinfo {pages} {036708} (\bibinfo {year}
  {2007})}\BibitemShut {NoStop}%
\bibitem [{\citenamefont {Chen}\ \emph {et~al.}(1999)\citenamefont {Chen},
  \citenamefont {Lov{\'a}sz},\ and\ \citenamefont {Pak}}]{chen1999lifting}%
  \BibitemOpen
  \bibfield  {author} {\bibinfo {author} {\bibfnamefont {F.}~\bibnamefont
  {Chen}}, \bibinfo {author} {\bibfnamefont {L.}~\bibnamefont {Lov{\'a}sz}},\
  and\ \bibinfo {author} {\bibfnamefont {I.}~\bibnamefont {Pak}},\ }\bibfield
  {title} {\bibinfo {title} {{Lifting Markov chains to speed up mixing}},\ }in\
  \href@noop {} {\emph {\bibinfo {booktitle} {Proceedings of the thirty-first
  annual ACM symposium on Theory of computing}}}\ (\bibinfo {year} {1999})\
  pp.\ \bibinfo {pages} {275--281}\BibitemShut {NoStop}%
\bibitem [{\citenamefont {Diaconis}\ \emph {et~al.}(2000)\citenamefont
  {Diaconis}, \citenamefont {Holmes},\ and\ \citenamefont
  {Neal}}]{diaconis2000analysis}%
  \BibitemOpen
  \bibfield  {author} {\bibinfo {author} {\bibfnamefont {P.}~\bibnamefont
  {Diaconis}}, \bibinfo {author} {\bibfnamefont {S.}~\bibnamefont {Holmes}},\
  and\ \bibinfo {author} {\bibfnamefont {R.~M.}\ \bibnamefont {Neal}},\
  }\bibfield  {title} {\bibinfo {title} {{Analysis of a nonreversible Markov
  chain sampler}},\ }\href@noop {} {\bibfield  {journal} {\bibinfo  {journal}
  {Annals of Applied Probability}\ ,\ \bibinfo {pages} {726}} (\bibinfo {year}
  {2000})}\BibitemShut {NoStop}%
\bibitem [{\citenamefont {Bernard}\ \emph {et~al.}(2009)\citenamefont
  {Bernard}, \citenamefont {Krauth},\ and\ \citenamefont
  {Wilson}}]{Bernard2009}%
  \BibitemOpen
  \bibfield  {author} {\bibinfo {author} {\bibfnamefont {E.~P.}\ \bibnamefont
  {Bernard}}, \bibinfo {author} {\bibfnamefont {W.}~\bibnamefont {Krauth}},\
  and\ \bibinfo {author} {\bibfnamefont {D.}~\bibnamefont {Wilson}},\
  }\bibfield  {title} {\bibinfo {title} {{Event-chain Monte Carlo algorithms
  for hard-sphere systems}},\ }\href
  {https://doi.org/10.1103/PhysRevE.80.056704} {\bibfield  {journal} {\bibinfo
  {journal} {Physical Review E}\ }\textbf {\bibinfo {volume} {80}},\ \bibinfo
  {pages} {056704} (\bibinfo {year} {2009})}\BibitemShut {NoStop}%
\bibitem [{\citenamefont {Suwa}\ and\ \citenamefont
  {Todo}(2010)}]{Suwa2010markov}%
  \BibitemOpen
  \bibfield  {author} {\bibinfo {author} {\bibfnamefont {H.}~\bibnamefont
  {Suwa}}\ and\ \bibinfo {author} {\bibfnamefont {S.}~\bibnamefont {Todo}},\
  }\bibfield  {title} {\bibinfo {title} {{Markov Chain Monte Carlo Method
  without Detailed Balance}},\ }\href
  {https://doi.org/10.1103/PhysRevLett.105.120603} {\bibfield  {journal}
  {\bibinfo  {journal} {Physical Review Letters}\ }\textbf {\bibinfo {volume}
  {105}},\ \bibinfo {pages} {120603} (\bibinfo {year} {2010})}\BibitemShut
  {NoStop}%
\bibitem [{\citenamefont {Turitsyn}\ \emph {et~al.}(2011)\citenamefont
  {Turitsyn}, \citenamefont {Chertkov},\ and\ \citenamefont
  {Vucelja}}]{Turitsyn2011irreversible}%
  \BibitemOpen
  \bibfield  {author} {\bibinfo {author} {\bibfnamefont {K.~S.}\ \bibnamefont
  {Turitsyn}}, \bibinfo {author} {\bibfnamefont {M.}~\bibnamefont {Chertkov}},\
  and\ \bibinfo {author} {\bibfnamefont {M.}~\bibnamefont {Vucelja}},\
  }\bibfield  {title} {\bibinfo {title} {{Irreversible Monte Carlo algorithms
  for efficient sampling}},\ }\href
  {https://doi.org/10.1016/j.physd.2010.10.003} {\bibfield  {journal} {\bibinfo
   {journal} {Physica D: Nonlinear Phenomena}\ }\textbf {\bibinfo {volume}
  {240}},\ \bibinfo {pages} {410} (\bibinfo {year} {2011})}\BibitemShut
  {NoStop}%
\bibitem [{\citenamefont {Fernandes}\ and\ \citenamefont
  {Weigel}(2011)}]{Fernandes2011nonreversible}%
  \BibitemOpen
  \bibfield  {author} {\bibinfo {author} {\bibfnamefont {H.~C.}\ \bibnamefont
  {Fernandes}}\ and\ \bibinfo {author} {\bibfnamefont {M.}~\bibnamefont
  {Weigel}},\ }\bibfield  {title} {\bibinfo {title} {{Non-reversible Monte
  Carlo simulations of spin models}},\ }\href
  {https://doi.org/10.1016/j.cpc.2010.11.017} {\bibfield  {journal} {\bibinfo
  {journal} {Computer Physics Communications}\ }\textbf {\bibinfo {volume}
  {182}},\ \bibinfo {pages} {1856} (\bibinfo {year} {2011})}\BibitemShut
  {NoStop}%
\bibitem [{\citenamefont {Engel}\ \emph {et~al.}(2013)\citenamefont {Engel},
  \citenamefont {Anderson}, \citenamefont {Glotzer}, \citenamefont {Isobe},
  \citenamefont {Bernard},\ and\ \citenamefont {Krauth}}]{Engel2013harddisk}%
  \BibitemOpen
  \bibfield  {author} {\bibinfo {author} {\bibfnamefont {M.}~\bibnamefont
  {Engel}}, \bibinfo {author} {\bibfnamefont {J.~A.}\ \bibnamefont {Anderson}},
  \bibinfo {author} {\bibfnamefont {S.~C.}\ \bibnamefont {Glotzer}}, \bibinfo
  {author} {\bibfnamefont {M.}~\bibnamefont {Isobe}}, \bibinfo {author}
  {\bibfnamefont {E.~P.}\ \bibnamefont {Bernard}},\ and\ \bibinfo {author}
  {\bibfnamefont {W.}~\bibnamefont {Krauth}},\ }\bibfield  {title} {\bibinfo
  {title} {{Hard-disk equation of state: First-order liquid-hexatic transition
  in two dimensions with three simulation methods}},\ }\href
  {https://doi.org/10.1103/PhysRevE.87.042134} {\bibfield  {journal} {\bibinfo
  {journal} {Physical Review E}\ }\textbf {\bibinfo {volume} {87}},\ \bibinfo
  {pages} {042134} (\bibinfo {year} {2013})}\BibitemShut {NoStop}%
\bibitem [{\citenamefont {Michel}\ \emph {et~al.}(2014)\citenamefont {Michel},
  \citenamefont {Kapfer},\ and\ \citenamefont
  {Krauth}}]{Michel2014generalized}%
  \BibitemOpen
  \bibfield  {author} {\bibinfo {author} {\bibfnamefont {M.}~\bibnamefont
  {Michel}}, \bibinfo {author} {\bibfnamefont {S.~C.}\ \bibnamefont {Kapfer}},\
  and\ \bibinfo {author} {\bibfnamefont {W.}~\bibnamefont {Krauth}},\
  }\bibfield  {title} {\bibinfo {title} {{Generalized event-chain Monte Carlo:
  Constructing rejection-free global-balance algorithms from infinitesimal
  steps}},\ }\href {https://doi.org/10.1063/1.4863991} {\bibfield  {journal}
  {\bibinfo  {journal} {The Journal of Chemical Physics}\ }\textbf {\bibinfo
  {volume} {140}},\ \bibinfo {pages} {054116} (\bibinfo {year}
  {2014})}\BibitemShut {NoStop}%
\bibitem [{\citenamefont {Nishikawa}\ \emph {et~al.}(2015)\citenamefont
  {Nishikawa}, \citenamefont {Michel}, \citenamefont {Krauth},\ and\
  \citenamefont {Hukushima}}]{nishikawa2015event}%
  \BibitemOpen
  \bibfield  {author} {\bibinfo {author} {\bibfnamefont {Y.}~\bibnamefont
  {Nishikawa}}, \bibinfo {author} {\bibfnamefont {M.}~\bibnamefont {Michel}},
  \bibinfo {author} {\bibfnamefont {W.}~\bibnamefont {Krauth}},\ and\ \bibinfo
  {author} {\bibfnamefont {K.}~\bibnamefont {Hukushima}},\ }\bibfield  {title}
  {\bibinfo {title} {{Event-chain algorithm for the Heisenberg model: Evidence
  for z {$\simeq$} 1 dynamic scaling}},\ }\href
  {https://doi.org/10.1103/PhysRevE.92.063306} {\bibfield  {journal} {\bibinfo
  {journal} {Physical Review E}\ }\textbf {\bibinfo {volume} {92}},\ \bibinfo
  {pages} {063306} (\bibinfo {year} {2015})}\BibitemShut {NoStop}%
\bibitem [{\citenamefont {Isobe}\ and\ \citenamefont
  {Krauth}(2015)}]{Isobe2015hardsphere}%
  \BibitemOpen
  \bibfield  {author} {\bibinfo {author} {\bibfnamefont {M.}~\bibnamefont
  {Isobe}}\ and\ \bibinfo {author} {\bibfnamefont {W.}~\bibnamefont {Krauth}},\
  }\bibfield  {title} {\bibinfo {title} {{Hard-sphere melting and
  crystallization with event-chain Monte Carlo}},\ }\href
  {http://dx.doi.org/10.1063/1.4929529} {\bibfield  {journal} {\bibinfo
  {journal} {Journal of Chemical Physics}\ }\textbf {\bibinfo {volume} {143}}
  (\bibinfo {year} {2015})}\BibitemShut {NoStop}%
\bibitem [{\citenamefont {Lei}\ and\ \citenamefont
  {Krauth}(2018)}]{Lei2018irreversible}%
  \BibitemOpen
  \bibfield  {author} {\bibinfo {author} {\bibfnamefont {Z.}~\bibnamefont
  {Lei}}\ and\ \bibinfo {author} {\bibfnamefont {W.}~\bibnamefont {Krauth}},\
  }\bibfield  {title} {\bibinfo {title} {{Irreversible Markov chains in spin
  models: Topological excitations}},\ }\href
  {https://doi.org/10.1209/0295-5075/121/10008} {\bibfield  {journal} {\bibinfo
   {journal} {EPL (Europhysics Letters)}\ }\textbf {\bibinfo {volume} {121}},\
  \bibinfo {pages} {10008} (\bibinfo {year} {2018})}\BibitemShut {NoStop}%
\bibitem [{\citenamefont {Lei}\ \emph {et~al.}(2019)\citenamefont {Lei},
  \citenamefont {Krauth},\ and\ \citenamefont {Maggs}}]{Lei2019eventchain}%
  \BibitemOpen
  \bibfield  {author} {\bibinfo {author} {\bibfnamefont {Z.}~\bibnamefont
  {Lei}}, \bibinfo {author} {\bibfnamefont {W.}~\bibnamefont {Krauth}},\ and\
  \bibinfo {author} {\bibfnamefont {A.~C.}\ \bibnamefont {Maggs}},\ }\bibfield
  {title} {\bibinfo {title} {{Event-chain Monte Carlo with factor fields}},\
  }\href {https://doi.org/10.1103/PhysRevE.99.043301} {\bibfield  {journal}
  {\bibinfo  {journal} {Physical Review E}\ }\textbf {\bibinfo {volume} {99}},\
  \bibinfo {pages} {043301} (\bibinfo {year} {2019})}\BibitemShut {NoStop}%
\bibitem [{\citenamefont {Klement}\ and\ \citenamefont
  {Engel}(2019)}]{Klement2019efficient}%
  \BibitemOpen
  \bibfield  {author} {\bibinfo {author} {\bibfnamefont {M.}~\bibnamefont
  {Klement}}\ and\ \bibinfo {author} {\bibfnamefont {M.}~\bibnamefont
  {Engel}},\ }\bibfield  {title} {\bibinfo {title} {{Efficient equilibration of
  hard spheres with Newtonian event chains}},\ }\href
  {https://doi.org/10.1063/1.5090882} {\bibfield  {journal} {\bibinfo
  {journal} {The Journal of Chemical Physics}\ }\textbf {\bibinfo {volume}
  {150}},\ \bibinfo {pages} {174108} (\bibinfo {year} {2019})}\BibitemShut
  {NoStop}%
\bibitem [{\citenamefont {Maggs}(2024)}]{Maggs2024nonreversible}%
  \BibitemOpen
  \bibfield  {author} {\bibinfo {author} {\bibfnamefont {A.~C.}\ \bibnamefont
  {Maggs}},\ }\bibfield  {title} {\bibinfo {title} {{Non-reversible Monte
  Carlo: An example of “true” self-repelling motion}},\ }\href
  {https://doi.org/10.1209/0295-5075/ad64ff} {\bibfield  {journal} {\bibinfo
  {journal} {Europhysics Letters}\ }\textbf {\bibinfo {volume} {147}},\
  \bibinfo {pages} {21001} (\bibinfo {year} {2024})}\BibitemShut {NoStop}%
\bibitem [{\citenamefont {Li}\ \emph {et~al.}(2022)\citenamefont {Li},
  \citenamefont {Nishikawa}, \citenamefont {H{\"{o}}llmer}, \citenamefont
  {Carillo}, \citenamefont {Maggs},\ and\ \citenamefont
  {Krauth}}]{Li2022harddisk}%
  \BibitemOpen
  \bibfield  {author} {\bibinfo {author} {\bibfnamefont {B.}~\bibnamefont
  {Li}}, \bibinfo {author} {\bibfnamefont {Y.}~\bibnamefont {Nishikawa}},
  \bibinfo {author} {\bibfnamefont {P.}~\bibnamefont {H{\"{o}}llmer}}, \bibinfo
  {author} {\bibfnamefont {L.}~\bibnamefont {Carillo}}, \bibinfo {author}
  {\bibfnamefont {A.~C.}\ \bibnamefont {Maggs}},\ and\ \bibinfo {author}
  {\bibfnamefont {W.}~\bibnamefont {Krauth}},\ }\bibfield  {title} {\bibinfo
  {title} {{Hard-disk pressure computations—a historic perspective}},\ }\href
  {https://doi.org/10.1063/5.0126437} {\bibfield  {journal} {\bibinfo
  {journal} {The Journal of Chemical Physics}\ }\textbf {\bibinfo {volume}
  {157}},\ \bibinfo {pages} {234111} (\bibinfo {year} {2022})}\BibitemShut
  {NoStop}%
\bibitem [{\citenamefont {Ghimenti}\ \emph {et~al.}(2024)\citenamefont
  {Ghimenti}, \citenamefont {Berthier},\ and\ \citenamefont {van
  Wijland}}]{Ghimenti2024irreversible}%
  \BibitemOpen
  \bibfield  {author} {\bibinfo {author} {\bibfnamefont {F.}~\bibnamefont
  {Ghimenti}}, \bibinfo {author} {\bibfnamefont {L.}~\bibnamefont {Berthier}},\
  and\ \bibinfo {author} {\bibfnamefont {F.}~\bibnamefont {van Wijland}},\
  }\bibfield  {title} {\bibinfo {title} {{Irreversible Monte Carlo Algorithms
  for Hard Disk Glasses: From Event-Chain to Collective Swaps}},\ }\href
  {https://doi.org/10.1103/PhysRevLett.133.028202} {\bibfield  {journal}
  {\bibinfo  {journal} {Physical Review Letters}\ }\textbf {\bibinfo {volume}
  {133}},\ \bibinfo {pages} {028202} (\bibinfo {year} {2024})}\BibitemShut
  {NoStop}%
\bibitem [{\citenamefont {Berthier}\ \emph {et~al.}(2024)\citenamefont
  {Berthier}, \citenamefont {Ghimenti},\ and\ \citenamefont {van
  Wijland}}]{Berthier2024MonteCarlo}%
  \BibitemOpen
  \bibfield  {author} {\bibinfo {author} {\bibfnamefont {L.}~\bibnamefont
  {Berthier}}, \bibinfo {author} {\bibfnamefont {F.}~\bibnamefont {Ghimenti}},\
  and\ \bibinfo {author} {\bibfnamefont {F.}~\bibnamefont {van Wijland}},\
  }\bibfield  {title} {\bibinfo {title} {{Monte Carlo simulations of
  glass-forming liquids beyond Metropolis}},\ }\href
  {https://doi.org/10.1063/5.0225978} {\bibfield  {journal} {\bibinfo
  {journal} {The Journal of Chemical Physics}\ }\textbf {\bibinfo {volume}
  {161}},\ \bibinfo {pages} {114105} (\bibinfo {year} {2024})}\BibitemShut
  {NoStop}%
\bibitem [{\citenamefont {Nishikawa}\ \emph {et~al.}(2025)\citenamefont
  {Nishikawa}, \citenamefont {Ghimenti}, \citenamefont {Berthier},\ and\
  \citenamefont {van Wijland}}]{Nishikawa2025irreversible}%
  \BibitemOpen
  \bibfield  {author} {\bibinfo {author} {\bibfnamefont {Y.}~\bibnamefont
  {Nishikawa}}, \bibinfo {author} {\bibfnamefont {F.}~\bibnamefont {Ghimenti}},
  \bibinfo {author} {\bibfnamefont {L.}~\bibnamefont {Berthier}},\ and\
  \bibinfo {author} {\bibfnamefont {F.}~\bibnamefont {van Wijland}},\
  }\bibfield  {title} {\bibinfo {title} {{Irreversible swap algorithms for soft
  sphere glasses}},\ }\href {https://doi.org/10.1103/PhysRevE.111.045416}
  {\bibfield  {journal} {\bibinfo  {journal} {Physical Review E}\ }\textbf
  {\bibinfo {volume} {111}},\ \bibinfo {pages} {045416} (\bibinfo {year}
  {2025})}\BibitemShut {NoStop}%
\bibitem [{\citenamefont {Sakai}\ and\ \citenamefont
  {Hukushima}(2016{\natexlab{b}})}]{Sakai2016irreversible}%
  \BibitemOpen
  \bibfield  {author} {\bibinfo {author} {\bibfnamefont {Y.}~\bibnamefont
  {Sakai}}\ and\ \bibinfo {author} {\bibfnamefont {K.}~\bibnamefont
  {Hukushima}},\ }\bibfield  {title} {\bibinfo {title} {Irreversible simulated
  tempering},\ }\href {https://doi.org/10.7566/JPSJ.85.104002} {\bibfield
  {journal} {\bibinfo  {journal} {Journal of the Physical Society of Japan}\
  }\textbf {\bibinfo {volume} {85}},\ \bibinfo {pages} {104002} (\bibinfo
  {year} {2016}{\natexlab{b}})}\BibitemShut {NoStop}%
\bibitem [{\citenamefont {Faizi}\ \emph {et~al.}(2020)\citenamefont {Faizi},
  \citenamefont {Buigues}, \citenamefont {Deligiannidis},\ and\ \citenamefont
  {Rosta}}]{Faizi2020simulated}%
  \BibitemOpen
  \bibfield  {author} {\bibinfo {author} {\bibfnamefont {F.}~\bibnamefont
  {Faizi}}, \bibinfo {author} {\bibfnamefont {P.~J.}\ \bibnamefont {Buigues}},
  \bibinfo {author} {\bibfnamefont {G.}~\bibnamefont {Deligiannidis}},\ and\
  \bibinfo {author} {\bibfnamefont {E.}~\bibnamefont {Rosta}},\ }\bibfield
  {title} {\bibinfo {title} {Simulated tempering with irreversible gibbs
  sampling techniques},\ }\href {https://doi.org/10.1063/5.0025775} {\bibfield
  {journal} {\bibinfo  {journal} {The Journal of Chemical Physics}\ }\textbf
  {\bibinfo {volume} {153}},\ \bibinfo {pages} {214111} (\bibinfo {year}
  {2020})}\BibitemShut {NoStop}%
\bibitem [{\citenamefont {Zhou}\ and\ \citenamefont {Bhatt}(2005)}]{Zhou2005}%
  \BibitemOpen
  \bibfield  {author} {\bibinfo {author} {\bibfnamefont {C.}~\bibnamefont
  {Zhou}}\ and\ \bibinfo {author} {\bibfnamefont {R.~N.}\ \bibnamefont
  {Bhatt}},\ }\bibfield  {title} {\bibinfo {title} {{Understanding and
  improving the Wang-Landau algorithm}},\ }\href
  {https://doi.org/10.1103/PhysRevE.72.025701} {\bibfield  {journal} {\bibinfo
  {journal} {Physical Review E}\ }\textbf {\bibinfo {volume} {72}},\ \bibinfo
  {pages} {025701} (\bibinfo {year} {2005})}\BibitemShut {NoStop}%
\bibitem [{\citenamefont {Fort}\ \emph {et~al.}(2015)\citenamefont {Fort},
  \citenamefont {Jourdain}, \citenamefont {Kuhn}, \citenamefont
  {Leli{\`{e}}vre},\ and\ \citenamefont {Stoltz}}]{Fort2015}%
  \BibitemOpen
  \bibfield  {author} {\bibinfo {author} {\bibfnamefont {G.}~\bibnamefont
  {Fort}}, \bibinfo {author} {\bibfnamefont {B.}~\bibnamefont {Jourdain}},
  \bibinfo {author} {\bibfnamefont {E.}~\bibnamefont {Kuhn}}, \bibinfo {author}
  {\bibfnamefont {T.}~\bibnamefont {Leli{\`{e}}vre}},\ and\ \bibinfo {author}
  {\bibfnamefont {G.}~\bibnamefont {Stoltz}},\ }\bibfield  {title} {\bibinfo
  {title} {{Convergence of the Wang-Landau algorithm}},\ }\href
  {https://doi.org/10.1090/S0025-5718-2015-02952-4} {\bibfield  {journal}
  {\bibinfo  {journal} {Mathematics of Computation}\ }\textbf {\bibinfo
  {volume} {84}},\ \bibinfo {pages} {2297} (\bibinfo {year}
  {2015})}\BibitemShut {NoStop}%
\bibitem [{\citenamefont {Berg}\ and\ \citenamefont
  {Neuhaus}(1991)}]{Berg1991}%
  \BibitemOpen
  \bibfield  {author} {\bibinfo {author} {\bibfnamefont {B.~A.}\ \bibnamefont
  {Berg}}\ and\ \bibinfo {author} {\bibfnamefont {T.}~\bibnamefont {Neuhaus}},\
  }\bibfield  {title} {\bibinfo {title} {{Multicanonical algorithms for first
  order phase transitions}},\ }\href
  {https://doi.org/10.1016/0370-2693(91)91256-U} {\bibfield  {journal}
  {\bibinfo  {journal} {Physics Letters B}\ }\textbf {\bibinfo {volume}
  {267}},\ \bibinfo {pages} {249} (\bibinfo {year} {1991})}\BibitemShut
  {NoStop}%
\bibitem [{\citenamefont {Berg}\ and\ \citenamefont
  {Neuhaus}(1992)}]{Berg1992}%
  \BibitemOpen
  \bibfield  {author} {\bibinfo {author} {\bibfnamefont {B.~A.}\ \bibnamefont
  {Berg}}\ and\ \bibinfo {author} {\bibfnamefont {T.}~\bibnamefont {Neuhaus}},\
  }\bibfield  {title} {\bibinfo {title} {{Multicanonical ensemble: A new
  approach to simulate first-order phase transitions}},\ }\href
  {https://doi.org/10.1103/PhysRevLett.68.9} {\bibfield  {journal} {\bibinfo
  {journal} {Physical Review Letters}\ }\textbf {\bibinfo {volume} {68}},\
  \bibinfo {pages} {9} (\bibinfo {year} {1992})}\BibitemShut {NoStop}%
\bibitem [{\citenamefont {Hansmann}\ \emph {et~al.}(1993)\citenamefont
  {Hansmann}, \citenamefont {Berg},\ and\ \citenamefont
  {Neuhaus}}]{Hansmann1993MultimagneticalSimulations}%
  \BibitemOpen
  \bibfield  {author} {\bibinfo {author} {\bibfnamefont {U.}~\bibnamefont
  {Hansmann}}, \bibinfo {author} {\bibfnamefont {B.~A.}\ \bibnamefont {Berg}},\
  and\ \bibinfo {author} {\bibfnamefont {T.}~\bibnamefont {Neuhaus}},\
  }\bibfield  {title} {\bibinfo {title} {Multimagnetical simulations},\ }in\
  \href@noop {} {\emph {\bibinfo {booktitle} {Computer Simulation Studies in
  Condensed-Matter Physics V}}},\ \bibinfo {editor} {edited by\ \bibinfo
  {editor} {\bibfnamefont {D.~P.}\ \bibnamefont {Landau}}, \bibinfo {editor}
  {\bibfnamefont {K.~K.}\ \bibnamefont {Mon}},\ and\ \bibinfo {editor}
  {\bibfnamefont {H.-B.}\ \bibnamefont {Sch{\"u}ttler}}}\ (\bibinfo
  {publisher} {Springer Berlin Heidelberg},\ \bibinfo {address} {Berlin,
  Heidelberg},\ \bibinfo {year} {1993})\ pp.\ \bibinfo {pages}
  {145--149}\BibitemShut {NoStop}%
\bibitem [{\citenamefont {Zhou}\ \emph {et~al.}(2006)\citenamefont {Zhou},
  \citenamefont {Schulthess}, \citenamefont {Torbr{\"{u}}gge},\ and\
  \citenamefont {Landau}}]{Zhou2006wanglandau}%
  \BibitemOpen
  \bibfield  {author} {\bibinfo {author} {\bibfnamefont {C.}~\bibnamefont
  {Zhou}}, \bibinfo {author} {\bibfnamefont {T.}~\bibnamefont {Schulthess}},
  \bibinfo {author} {\bibfnamefont {S.}~\bibnamefont {Torbr{\"{u}}gge}},\ and\
  \bibinfo {author} {\bibfnamefont {D.~P.}\ \bibnamefont {Landau}},\ }\bibfield
   {title} {\bibinfo {title} {{Wang-Landau Algorithm for Continuous Models and
  Joint Density of States}},\ }\href
  {https://doi.org/10.1103/PhysRevLett.96.120201} {\bibfield  {journal}
  {\bibinfo  {journal} {Physical Review Letters}\ }\textbf {\bibinfo {volume}
  {96}},\ \bibinfo {pages} {120201} (\bibinfo {year} {2006})}\BibitemShut
  {NoStop}%
\bibitem [{\citenamefont {Tsai}\ \emph {et~al.}(2009)\citenamefont {Tsai},
  \citenamefont {Wang},\ and\ \citenamefont {Landau}}]{Tsai2009}%
  \BibitemOpen
  \bibfield  {author} {\bibinfo {author} {\bibfnamefont {S.-H.}\ \bibnamefont
  {Tsai}}, \bibinfo {author} {\bibfnamefont {F.}~\bibnamefont {Wang}},\ and\
  \bibinfo {author} {\bibfnamefont {D.~P.}\ \bibnamefont {Landau}},\ }\bibfield
   {title} {\bibinfo {title} {{Two-Dimensional Wang-Landau Sampling of an
  Asymmetric Ising Model}},\ }\href {https://doi.org/10.1142/S0129183109014424}
  {\bibfield  {journal} {\bibinfo  {journal} {International Journal of Modern
  Physics C}\ }\textbf {\bibinfo {volume} {20}},\ \bibinfo {pages} {1357}
  (\bibinfo {year} {2009})}\BibitemShut {NoStop}%
\bibitem [{\citenamefont {Atchad{\'e}}\ and\ \citenamefont
  {Liu}(2010)}]{Atchade2010wanglandau}%
  \BibitemOpen
  \bibfield  {author} {\bibinfo {author} {\bibfnamefont {Y.~F.}\ \bibnamefont
  {Atchad{\'e}}}\ and\ \bibinfo {author} {\bibfnamefont {J.~S.}\ \bibnamefont
  {Liu}},\ }\bibfield  {title} {\bibinfo {title} {{The Wang-Landau algorithm in
  general state spaces: Applications and convergence analysis}},\ }\href
  {http://www.jstor.org/stable/24308988} {\bibfield  {journal} {\bibinfo
  {journal} {Statistica Sinica}\ }\textbf {\bibinfo {volume} {20}},\ \bibinfo
  {pages} {209} (\bibinfo {year} {2010})}\BibitemShut {NoStop}%
\bibitem [{\citenamefont {Nishikawa}\ and\ \citenamefont
  {Hukushima}(2020)}]{Nishikawa2020lattice}%
  \BibitemOpen
  \bibfield  {author} {\bibinfo {author} {\bibfnamefont {Y.}~\bibnamefont
  {Nishikawa}}\ and\ \bibinfo {author} {\bibfnamefont {K.}~\bibnamefont
  {Hukushima}},\ }\bibfield  {title} {\bibinfo {title} {{Lattice Glass Model in
  Three Spatial Dimensions}},\ }\href
  {https://doi.org/10.1103/PhysRevLett.125.065501} {\bibfield  {journal}
  {\bibinfo  {journal} {Physical Review Letters}\ }\textbf {\bibinfo {volume}
  {125}},\ \bibinfo {pages} {065501} (\bibinfo {year} {2020})}\BibitemShut
  {NoStop}%
\bibitem [{\citenamefont {Edwards}\ and\ \citenamefont
  {Anderson}(1975)}]{Edwards1975theory}%
  \BibitemOpen
  \bibfield  {author} {\bibinfo {author} {\bibfnamefont {S.~F.}\ \bibnamefont
  {Edwards}}\ and\ \bibinfo {author} {\bibfnamefont {P.~W.}\ \bibnamefont
  {Anderson}},\ }\bibfield  {title} {\bibinfo {title} {{Theory of spin
  glasses}},\ }\href {https://doi.org/10.1088/0305-4608/5/5/017} {\bibfield
  {journal} {\bibinfo  {journal} {Journal of Physics F: Metal Physics}\
  }\textbf {\bibinfo {volume} {5}},\ \bibinfo {pages} {965} (\bibinfo {year}
  {1975})}\BibitemShut {NoStop}%
\bibitem [{\citenamefont {Katzgraber}\ \emph {et~al.}(2006)\citenamefont
  {Katzgraber}, \citenamefont {K{\"{o}}rner},\ and\ \citenamefont
  {Young}}]{Katzgraber2006universality}%
  \BibitemOpen
  \bibfield  {author} {\bibinfo {author} {\bibfnamefont {H.~G.}\ \bibnamefont
  {Katzgraber}}, \bibinfo {author} {\bibfnamefont {M.}~\bibnamefont
  {K{\"{o}}rner}},\ and\ \bibinfo {author} {\bibfnamefont {A.~P.}\ \bibnamefont
  {Young}},\ }\bibfield  {title} {\bibinfo {title} {{Universality in
  three-dimensional Ising spin glasses: A Monte Carlo study}},\ }\href
  {https://doi.org/10.1103/PhysRevB.73.224432} {\bibfield  {journal} {\bibinfo
  {journal} {Physical Review B}\ }\textbf {\bibinfo {volume} {73}},\ \bibinfo
  {pages} {224432} (\bibinfo {year} {2006})}\BibitemShut {NoStop}%
\bibitem [{\citenamefont {Katzgraber}\ and\ \citenamefont
  {Campbell}(2005)}]{Katzgraber2005dynamical}%
  \BibitemOpen
  \bibfield  {author} {\bibinfo {author} {\bibfnamefont {H.~G.}\ \bibnamefont
  {Katzgraber}}\ and\ \bibinfo {author} {\bibfnamefont {I.~A.}\ \bibnamefont
  {Campbell}},\ }\bibfield  {title} {\bibinfo {title} {{Dynamical scaling in
  Ising and vector spin glasses}},\ }\href
  {https://doi.org/10.1103/PhysRevB.72.014462} {\bibfield  {journal} {\bibinfo
  {journal} {Physical Review B}\ }\textbf {\bibinfo {volume} {72}},\ \bibinfo
  {pages} {014462} (\bibinfo {year} {2005})}\BibitemShut {NoStop}%
\bibitem [{\citenamefont {Ferrenberg}\ and\ \citenamefont
  {Swendsen}(1988)}]{Ferrenberg1988new}%
  \BibitemOpen
  \bibfield  {author} {\bibinfo {author} {\bibfnamefont {A.~M.}\ \bibnamefont
  {Ferrenberg}}\ and\ \bibinfo {author} {\bibfnamefont {R.~H.}\ \bibnamefont
  {Swendsen}},\ }\bibfield  {title} {\bibinfo {title} {{New Monte Carlo
  technique for studying phase transitions}},\ }\href
  {https://doi.org/10.1103/PhysRevLett.61.2635} {\bibfield  {journal} {\bibinfo
   {journal} {Physical Review Letters}\ }\textbf {\bibinfo {volume} {61}},\
  \bibinfo {pages} {2635} (\bibinfo {year} {1988})}\BibitemShut {NoStop}%
\bibitem [{\citenamefont {Wang}(2015)}]{Wang2015measuring}%
  \BibitemOpen
  \bibfield  {author} {\bibinfo {author} {\bibfnamefont {W.}~\bibnamefont
  {Wang}},\ }\bibfield  {title} {\bibinfo {title} {{Measuring free energy in
  spin-lattice models using parallel tempering Monte Carlo}},\ }\href
  {https://doi.org/10.1103/PhysRevE.91.053303} {\bibfield  {journal} {\bibinfo
  {journal} {Physical Review E}\ }\textbf {\bibinfo {volume} {91}},\ \bibinfo
  {pages} {053303} (\bibinfo {year} {2015})}\BibitemShut {NoStop}%
\bibitem [{\citenamefont {Yucesoy}\ \emph {et~al.}(2013)\citenamefont
  {Yucesoy}, \citenamefont {Machta},\ and\ \citenamefont
  {Katzgraber}}]{Yucesoy2013correlations}%
  \BibitemOpen
  \bibfield  {author} {\bibinfo {author} {\bibfnamefont {B.}~\bibnamefont
  {Yucesoy}}, \bibinfo {author} {\bibfnamefont {J.}~\bibnamefont {Machta}},\
  and\ \bibinfo {author} {\bibfnamefont {H.~G.}\ \bibnamefont {Katzgraber}},\
  }\bibfield  {title} {\bibinfo {title} {{Correlations between the dynamics of
  parallel tempering and the free-energy landscape in spin glasses}},\ }\href
  {https://doi.org/10.1103/PhysRevE.87.012104} {\bibfield  {journal} {\bibinfo
  {journal} {Physical Review E}\ }\textbf {\bibinfo {volume} {87}},\ \bibinfo
  {pages} {012104} (\bibinfo {year} {2013})}\BibitemShut {NoStop}%
\end{thebibliography}%
\end{document}